\documentclass[a4paper,twocolumn,amsmath,amssymb,pra,footinbib,floatfix,superscriptaddress]{revtex4}
\usepackage[english]{babel}
\usepackage[utf8]{inputenc}

\usepackage{graphicx}
\usepackage{dcolumn}
\usepackage{bm}
\usepackage{multirow}
\usepackage{mathdots}
\usepackage{amssymb}
\usepackage{amsmath}
\usepackage{hyperref}
\usepackage{color}
\usepackage{soul}
\usepackage{physics}
\usepackage{MnSymbol}
\usepackage{float}
\usepackage{mathtools}

\definecolor{red}{rgb}{0.7,0,0}
\definecolor{green}{rgb}{0.,0.35,0.}
\definecolor{blue}{rgb}{0.2,0.2,0.7} 
\definecolor{black}{rgb}{0.15,0.15,.15}
\def\mel#1#2#3{\langle #1 | #2 | #3 \rangle}
\def\ev#1#2{\mel{#2}{#1}{#2}} 
\def\CFT{\textrm{CFT}}

\def\({\left(}
\def\){\right)}
\def\<{\langle}
\def\>{\rangle}
\def\{{\lbrace}
\def\}{\rbrace}
\def\({\left(}
\def\){\right)}
\def\[{\left[}
\def\]{\right]}

\def\R{{\mathbb R}}

\def\eff{\text{eff}}
\def\scs{\text{scs}}

\DeclarePairedDelimiter\floor{\lfloor}{\rfloor}
\usepackage{tikz}
\usetikzlibrary{shapes.geometric,arrows}
\makeindex

\begin{document}

\title{Piercing  the rainbow: entanglement on an  inhomogeneous spin chain with a defect}

\author{Nadir Samos Sáenz de Buruaga}
\affiliation{Instituto de Física Teórica UAM/CSIC, Universidad
  Autónoma de Madrid, Cantoblanco, Madrid, Spain}

\author{Silvia N. Santalla}
\affiliation{Dto. Física \&\ GISC, Universidad Carlos III de
  Madrid, Spain}

\author{Javier Rodríguez-Laguna}
\affiliation{Dto. Física Fundamental, Universidad Nacional de
  Educación a Distancia (UNED), Madrid, Spain}

\author{Germán Sierra}
\affiliation{Instituto de Física Teórica UAM/CSIC, Universidad
  Autónoma de Madrid, Cantoblanco, Madrid, Spain}

\begin{abstract}
The {\em rainbow state} denotes a set of valence bond states organized
concentrically around the center of a spin 1/2 chain. It is the ground
state of an inhomogeneous XX Hamiltonian and presents maximal
violation of the area law of entanglement entropy.  Here, we add a
tunable exchange coupling constant at the center, $\gamma$, and show
that it induces entanglement transitions of the ground state.  At very
strong inhomogeneity, the rainbow state survives for $0 \leq \gamma
\leq 1$, while outside that region the ground state is a product of
dimers.  In the weak inhomogeneity regime the entanglement entropy
satisfies a volume law, derived from CFT in curved spacetime, with an
effective central charge that depends on the inhomogeneity parameter
and $\gamma$. In all regimes we have found that the entanglement
properties are invariant under the transformation $\gamma
\longleftrightarrow 1 - \gamma$, whose fixed point $\gamma =
\frac{1}{2}$ corresponds to the usual rainbow model.  Finally, we
study the robustness of non trivial topological phases in the presence
of the defect.
\end{abstract}

\date{December 23, 2019}

\maketitle


\section{Introduction} 
\label{sec:intro}

Entanglement provides a very useful connecting thread through
condensed matter physics, quantum optics and quantum field theory,
towards a unified field of {\em quantum matter}
\cite{Amico.08,Laflorencie.16}. One of the most relevant insights is
expressed through the {\em area law} \cite{Hastings2006,Wolf2008}: the entanglement entropy of a
block within the ground state (GS) of a local quantum system is, in
general terms, proportional to the measure of its boundary
\cite{Sredniki.93,Eisert.10}. Interestingly, the GS of a (1+1)D
conformal field is an exception, and the entropy of a block is
generically proportional to the {\em logarithm} of its volume, with a
prefactor which is proportional to the associated central charge
\cite{Holzhey.94,Vidal.03,Calabrese.04,Calabrese.09}. There are other
interesting exceptions, such as random systems
\cite{Refael.04,Refael.04b,Laflorencie.05,Fagotti.11,Ramirez.14,Ruggiero.16}. In
the strong inhomogeneity limit, the GS of many random systems can be
obtained via the Dasgupta-Ma procedure \cite{Dasgupta.80,Fisher.95},
which can be {\em engineered} to obtain a 1D GS with maximal
entanglement between its left and right halves, known as the {\em
  rainbow state}
\cite{Vitagliano.10,Ramirez.14b,Ramirez.15,Samos.19}. This violation
of the area law is very robust with respect to the presence of
disorder in the hoppings \cite{Laguna.16,Alba.19}.

A physical interpretation of the rainbow state can be provided by
noticing that the Dirac vacuum on a static (1+1)D metric of optical
type can be simulated on a free fermionic lattice with smoothly
varying hopping amplitudes \cite{Boada.11,Laguna.17}. Indeed, the
hopping amplitudes which characterize the rainbow state can be
understood as a (1+1)D anti-de Sitter (AdS) metric
\cite{Laguna.17b,Tonni.18,MacCormack.18}. Space is exponentially
stretched as we move away from the center, giving rise to a similar
exponential stretch of the entanglement entropies, transforming the
logarithmic law into a volume law. The weak inhomogeneity limit is
determined by a deformation of the conformal law and the strong
inhomogeneity limit is determined by the Dasgupta-Ma rule, and both
fit seamlessly.

Thus, it is relevant to ask whether the weak and the strong
inhomogeneity limits will match in all possible situations. We have
introduced a {\em defect} in the center of the rainbow system and
considered the entanglement structure as a function of the defect
intensity and the curvature. As we will show, both the Dasgupta-Ma and
the field theory approach that describes entanglement on a critical
chain with a defect \cite{Levine.04,Peschel.05,Igloi.09,Eisler.10} can
be extended to the curved case in the strong and weak inhomogeneity
regimes, respectively, providing a complete physical picture.

This article is organized as follows. Section \ref{sec:model}
discusses our model. The strong inhomogeneity limit, studied with the
Dasgupta-Ma RG, is described in detail in Sec. \ref{sec:strong}, while
Sec. \ref{sec:weak} considers the weak-inhomogeneity regime through a
perturbation of a conformal field theory. We characterize the
entanglement structure via the entropies and the entanglement
spectrum, Hamiltonian and contour. A duplication of the defect leads
the system to a symmetry protected topological (SPT) phase in
coexistence with a trivial dimerized phase, which are discussed in
Sec. \ref{sec:coexistence}. The article ends with a brief discussion
of our conclusions and proposals for further work in
Sec. \ref{sec:conclusions}.


\section{The model}
\label{sec:model}

Let us consider a fermionic open 1D free fermion system with an even
number of sites $N=2L$ whose Hamiltonian is defined as:

\begin{equation}
  H_L(h,\gamma)=
  -\frac{1}{2}\sum_{m=-L+1}^{L-1}
  J_m c^\dagger_{m-1/2} c_{m+1/2} + \text{h.c.},
  \label{eq:ham}
\end{equation}
where $c_n (c^\dagger_n)$, with $n=\pm\frac{1}{2},\pm\frac{3}{2}\dots\pm
(L-\frac{1}{2})$ are fermionic annihilation (creation) operators that obey the standard
anti-commutation relations. The hopping parameters $J_m$ are

\begin{equation}
J_m = \begin{cases}
  e^{-h|m|} & \text{if } m\neq 0, \\
  e^{-h\gamma} & m=0,\\
\end{cases}
\label{eq:hop_rainbow}
\end{equation}
where $h\geq 0$ is the inhomogeneity parameter, and $\gamma\in\R$
parametrizes the value of the central hopping that we shall interpret
as a {\em defect}. Notice that sites have half-integer indices,
while links have integer ones. The Hamiltonian presents spatial
inversion symmetry around the central bond: $J_m=J_{-m}$, which we
will label {\em bond-centered symmetry} (bcs) \cite{Samos.19}.

\begin{figure}
  \includegraphics[width=8cm]{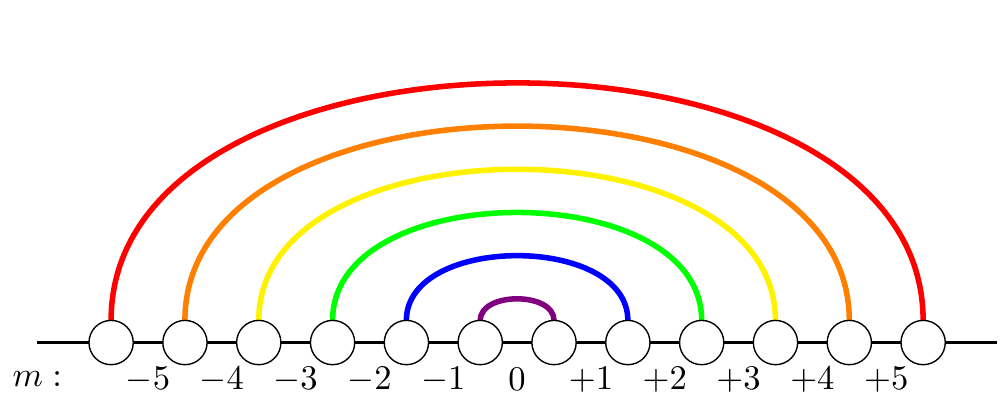}
  \caption{Illustrating the rainbow state, GS of $H_L(h,1/2)$ with
    $L=6$, for $h\gg 1$. Links are indexed by the integer $m$. The
    bonds are established between sites $n$ and $-n$ for $n\in \{\pm
    1/2,\cdots,\pm L/2-1\}$.}
  \label{fig:illust_rainbow}
\end{figure}

For $\gamma=1/2$ we recover the so-called rainbow Hamiltonian, whose
ground state is the {\em rainbow state}
\cite{Vitagliano.10,Ramirez.14b,Ramirez.15}. In the strong-coupling
regime ($h\gg 1$), the GS of $H_L(h,1/2)$ is a valence bond solid
(VBS) made of bonds connecting opposite sites of the chain, as is
illustrated in Fig. \ref{fig:illust_rainbow}. On the other hand, the
weak coupling limit ($h\ll 1$) is characterized by the free-fermion
conformal field theory (CFT) on a different space-time, provided by
the metric

\begin{equation}
ds^2=-e^{-2h|x|}dt^2 + dx^2,
\label{eq:metric}
\end{equation}
thus justifying our claim that the rainbow state, in the weak coupling
regime, corresponds to the anti de Sitter (AdS) Dirac vacuum. Using
CFT tools, it has been proved that the GS of Hamiltonian $H_L(h,1/2)$
presents linear entanglement for all $h$, with an {\em entropy per
  site} $S/N\approx h/6$ (von Neumann entanglement entropy)
\cite{Laguna.17b,Tonni.18}. We will discuss some of these properties
in detail in the corresponding sections, when considering how they are
modified by the defect.


\section{The strong inhomogeneity limit} 
\label{sec:strong}

When the inhomogeneity is large enough, it can be addressed through
renormalization group (RG) schemes. In particular we will use the
Dasgupta-Ma procedure, also known as strong-disorder renormalization
group (SDRG) \cite{Dasgupta.80}, that was originally created to
characterize random spin chains, but can be immediately extended to
fermionic chains via de Jordan-Wigner transformation
\cite{Ramirez.14}. At each step of the RG, four spins are considered:
the two spins ($s_i$ and $s_{i+1}$ linked by the strongest coupling
(highest absolute value of $J_i$) and their nearest neighbours,
$s_{i-1}$ and $s_{i+2}$. The two spins coupled by $J_i$ are integrated
out by forming a valence bond state (VBS) and the two remaining ones
are coupled with a new effective coupling constant that is obtained by
means of second order perturbation theory. For a free-fermionic chain
with a Hamiltonian such as \eqref{eq:ham}, the effective coupling
takes the expression

\begin{equation}
 \tilde J_i = -{J_{i-1}J_{i+1}\over J_i}, \qquad |J_i|\gg|J_{i\pm1}|.
\label{eq:sdrgsp}
\end{equation}  

The GS predicted by the SDRG is a valence-bond solid (VBS) i.e. a
tensor product of bonds:

\begin{equation}
  \ket{GS}=\prod_{k=1}^L
  (b^{\eta_k}_{i_k,j_k})^\dagger\ket{0},
  \label{eq:gs}
\end{equation}
where $\eta_k=\pm1$ is a phase given by Eq. \eqref{eq:sdrgsp}, $\ket{0}$ is the Fock vacuum and $b^{+}_{i,j}$ ($b^{-}_{i,j}$)
are bonding (anti-bonding) operators that create a fermionic
excitation joining sites $i$ and $j$.

\begin{equation}
  \(b^{\pm}_{ij}\)^\dagger=\frac{1}{\sqrt{2}}\(c^\dagger_{i}\pm c^\dagger_{j}\).
  \label{eq:bonding}
\end{equation}   

 They satisfy usual canonical anti-commutation relations. For our purposes, it is convenient to define the log-couplings
\cite{Laguna.16},

\begin{equation}
t_i=-\log\frac{|J_i|}{h},
\label{eq:logcoupling}
\end{equation}
where $h$ is the inhomogeneity parameter that is included for later
convenience. In this language, the two fermionic sites that are
integrated out are those connected by the {\em lowest} $t_i$ and the
effective coupling Eq. \eqref{eq:sdrgsp} is computed in additive
way:

\begin{equation}
\tilde t_i = t_{i-1}-t_i+t_{i+1}.
\label{eq:sdrg}
\end{equation}

Interestingly, Eq. \eqref{eq:sdrg} can be generalized for this type of
models: whenever a bond is established between sites $p$ and $q$ (with
$p+q$ even), the renormalized log-coupling is given by the {\em
  sum rule} \cite{Alba.19}:

\begin{equation}
\tilde t_{[p,q]} = \sum_{j=0}^{q-p-1} (-1)^j t_{p+j},
\label{eq:teorema}
\end{equation}
or, in other words: the renormalized log-coupling can be obtained
summing all log-coupling between the two extremes, with alternating
signs.

\bigskip

\begin{figure*}
  \includegraphics[width=16cm]{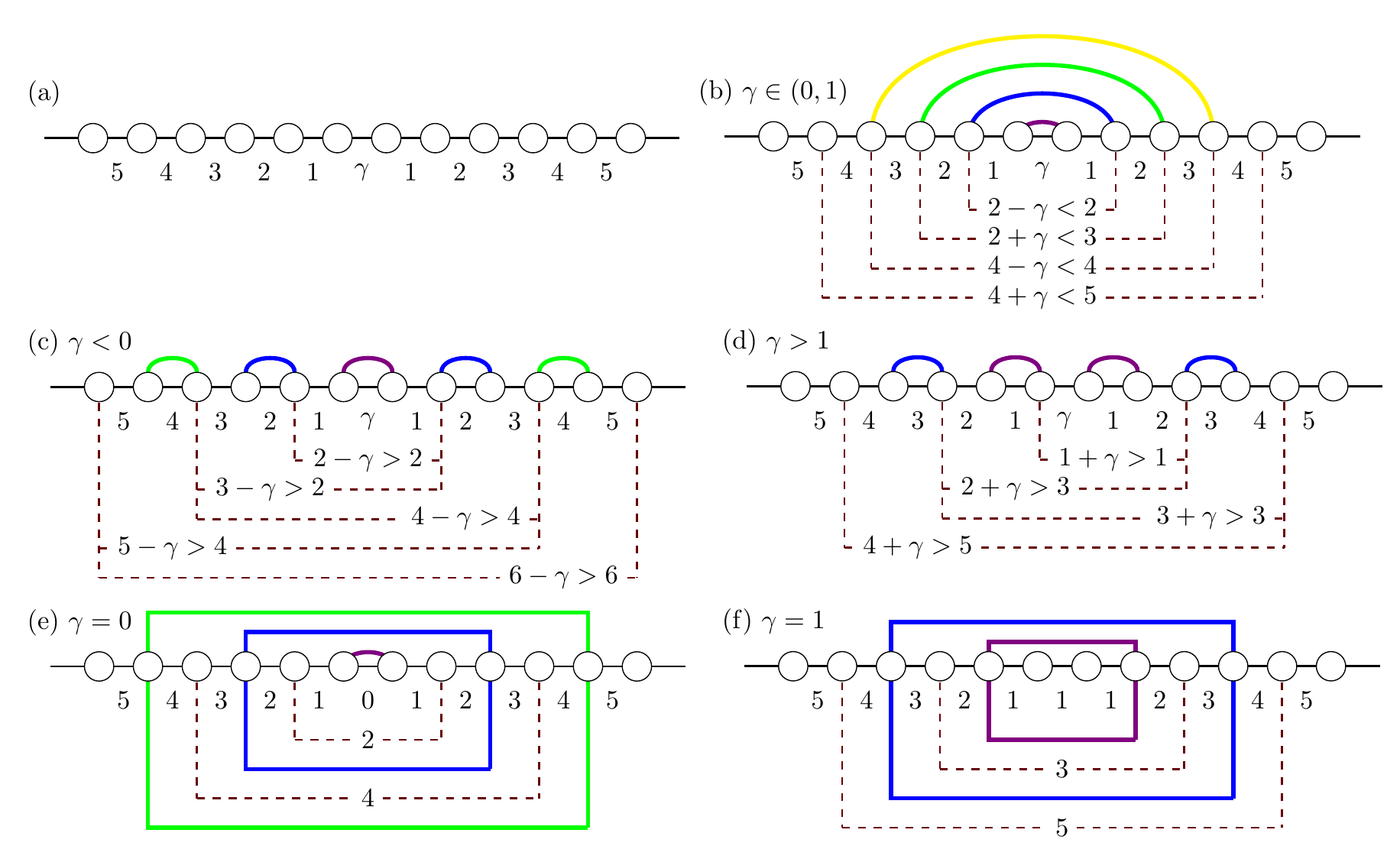}
  \caption{(a) Illustration of the rainbow chain with a central
    defect, showing the log-couplings on each link; (b) SDRG procedure
    in the $\gamma\in (0,1)$ case, leading to the {\em rainbow phase};
    (c) SDRG for the $\gamma<0$ case; (d) SDRG for the $\gamma>1$
    case, both leading to {\em dimerized phases}; (e) and (f) transition
    cases, where the SDRG approximation is not valid; the dashed boxes
    mark the {\em ties} between the couplings, which demand a
    different RG approach.}
  \label{fig:strong}
\end{figure*}

Let us consider the GS of Hamiltonian \eqref{eq:ham} under the light
of the SDRG for $h\gg 1$. For simplicity, we will only consider even
$L$ (the case of odd $L$ can be straightforwardly obtained), as a
function of the defect parameter, $\gamma$. The different phases will
be discussed along the panels of Fig. \ref{fig:strong}. In
Fig. \ref{fig:strong} (a) we have plotted the values of the
log-couplings, for later reference:
$\{\cdots,2,1,\gamma,1,2,\cdots\}$.

\begin{itemize}

\item {\em The rainbow phase:} $\gamma \in(0,1)$, see
  Fig. \ref{fig:strong} (b). The strongest link (lowest log-coupling)
  is the central one. Thus, a valence bond is established on top
  ($b^+_{-1/2,+1/2}$) and an effective log-coupling appears between its
  neighbors, of magnitude $\tilde t=2-\gamma<2$. Thus, the central link is
  again the strongest one, so we can put a valence-bond on top of it
  ($b^-_{-3/2,+3/2}$), leading to an effective log-coupling of
  magnitude $\tilde t=2+\gamma<3$. We can see that the procedure
  iterates, giving rise to the rainbow state.
  
\begin{equation}	
\label{eq:gsr}
\ket{GS}_{\gamma\in(0,1)}=
\prod_{i=0}^{L-1}\(b^{\eta_i}_{-i-1/2,i+1/2}\)^\dagger\ket{0},\quad \eta_i=(-1)^i.
\end{equation}

\item {\em The dimerized phase (I):} $\gamma<0$, see Fig. \ref{fig:strong}
  (c). The dominant interaction is again the central one, leading us
  to establish a valence bond on top. Yet, the renormalized
  log-coupling, $\tilde t=2-\gamma>2$ is not the strongest (lowest
  value) at the next SDRG iteration. On the other hand, we are led to
  establish two valence bonds on top of the links with log-couplings
  equal to 2, in any order. The renormalized central log-coupling
  after these two bonds have been established is $\tilde t=4-\gamma>4$
  (see Eq. \eqref{eq:teorema}), so we are led to the same situation,
  where the lateral links are always stronger than the central one,
  leading to a {\em dimerized state}. Yet, the last SDRG step leaves
  us with the two extreme sites of the chain, leading to a final bond
  connecting them. The ground state can be written as

\begin{equation}
  \ket{GS}_{\gamma<0}=
  (b^-_{-L+\frac{1}{2},L-\frac{1}{2}})^\dagger
  \prod_{i=-\frac{L}{2}+1}^{\frac{L}{2}-1}
  (b^+_{2i-\frac{1}{2},2i+\frac{1}{2}})^\dagger\ket{0}.
  \label{eq:gsdimer}	
\end{equation}

  Notice that the last bond is only present for even $L$, while it is
  absent for odd $L$.

\item {\em The dimerized phase (II):} $\gamma>1$, see
  Fig. \ref{fig:strong} (d). In this case, the dominant interaction is
  {\em not} the central one, but their neighbors, with $t_{\pm
    1}=1$. Hence, we must establish first these two valence bonds,
  leading to a renormalized log-coupling between their extremes of
  $\tilde t=2+\gamma>3$ (see Eq. \eqref{eq:teorema}). Thus, we have
  the same situation, in which the central link is not the
  strongest. In this case, no long-range bond is established at the
  end of the procedure, and we obtain

\begin{equation}
  \ket{GS}_{\gamma>1}=\prod_{i=-\frac{L-1}{2}}^{\frac{L-1}{2}}
  (b^+_{2i-\frac{1}{2},2i+\frac{1}{2}})^\dagger\ket{0}.
  \label{eq:gsdimer1}
\end{equation}

 Both dimerized phases are related to the two phases of the
  Su-Schrieffer-Heeger (SSH) model \cite{Su.79,Heeger.88}.

\item {\em The transition points: $\gamma=1$ and $\gamma=0$}, see
  Fig. \ref{fig:strong} (e) and (f). Let us start with $\gamma=1$
  (Fig. \ref{fig:strong} (e)). The first SDRG step fails, because the
  strongest coupling is not unique. On the other hand, we obtain a
  {\em triple tie} in the three central links, with $t_{0,\pm1}=1$. In
  Appendix \ref{sec:rgextension} we have developed an extension of the
  SDRG for the free-fermion model when a block with $2\ell$ sites is
  integrated out, yielding the renormalized log-coupling given by the
  sum rule, Eq. \eqref{eq:teorema}. Thus, the renormalized
  log-coupling between sites $-5/2$ and $+5/2$ is $\tilde t=3$,
  leading to a new triple tie, which propagates further along the
  chain. For $\gamma=0$, on the other hand, the strongest link is the
  central one, thus receiving a valence bond. But, on the next SDRG
  step, we can see that the effective central log-coupling is $\tilde
  t=2$, equal to its neighbors in a new {\em triple tie}, forcing us
  to recourse to the extended SDRG. From that moment on, all SDRG
  steps lead to triple ties. The GS can be written exactly in these
  two cases (see details in Appendix \ref{sec:single_body_modes})

\begin{equation}
  \ket{GS}_{\gamma=0}= (b^-_{-L+\frac{1}{2},L-\frac{1}{2}})^\dagger
  \prod_{i=1}^{\frac{L}{4}} (d^{\eta_i}_{2i+\frac{1}{2}})^\dagger
  (b^+_{-\frac{1}{2},\frac{1}{2}})^\dagger\ket{0},
\label{eq:gs0}
\end{equation}

\begin{equation}
\ket{GS}_{\gamma=1}= \prod_{i=1}^{\frac{L}{2}}(d^{\eta_i}_{2i-\frac{1}{2}})^\dagger\ket{0},
\label{eq:gs1} 
\end{equation}
where $d^\pm_k$ is operators creating two particles on four fermionic
sites,

\begin{eqnarray}
\label{eq:bond4c}
&(d^{\eta_i}_i)^\dagger=(v^{\eta_i})^\dagger(u^{\eta_{i-1}})^\dagger\ket{0},
\quad \eta_i=(-1)^{i},	\\
&u^\pm_i= \frac{1}{\sqrt{5+\sqrt{5}}}
(c_{-i}\pm c_i)+\frac{1}{\sqrt{5-\sqrt{5}}}(c_{-i+1}\pm c_{i-1}),\\
&v^\pm_i= \frac{1}{\sqrt{5-\sqrt{5}}}(c_{-i}\pm c_i)+
\frac{1}{\sqrt{5+\sqrt{5}}}(c_{-i+1}\pm c_{i-1}).
\end{eqnarray}
\end{itemize}

The aforementioned description, along with the evidences presented in
the rest of this section allow us to claim that the rainbow system
with a defect presents two {\em entanglement transitions} \cite{Vasseur2018} in the
strong inhomogeneity regime.

\bigskip

It is worth to notice the existence of a symmetry between the cases
$\gamma\leq0$ and $\gamma\geq1$. Consider a system
$H_L(h,\gamma<1)$. After performing the first RG step, the new system
is described by the renormalized Hamiltonian $H_{L-1}(h,2-\gamma)$. If
we now subtract one from all the log-hoppings (or equivantly we divide
by $e^{-h}$ all the hoppings) the Hamiltonian becomes $e^h
H_{L-1}(h,1-\gamma)$, which describes a system of $N-2$ sites and a
defect with strength $1-\gamma$. Hence, the transformation

\begin{equation}
  \gamma\to \tilde\gamma=1-\gamma,
  \label{eq:symmetry}
\end{equation}
leaves the structure invariant (up to a global constant). Note that
this symmetry can be considered as a local strong-weak duality of the
defects, leaving the $\gamma=1/2$ point invariant.

\bigskip

\begin{figure}
  \begin{center}
    \includegraphics[width=8.8cm]{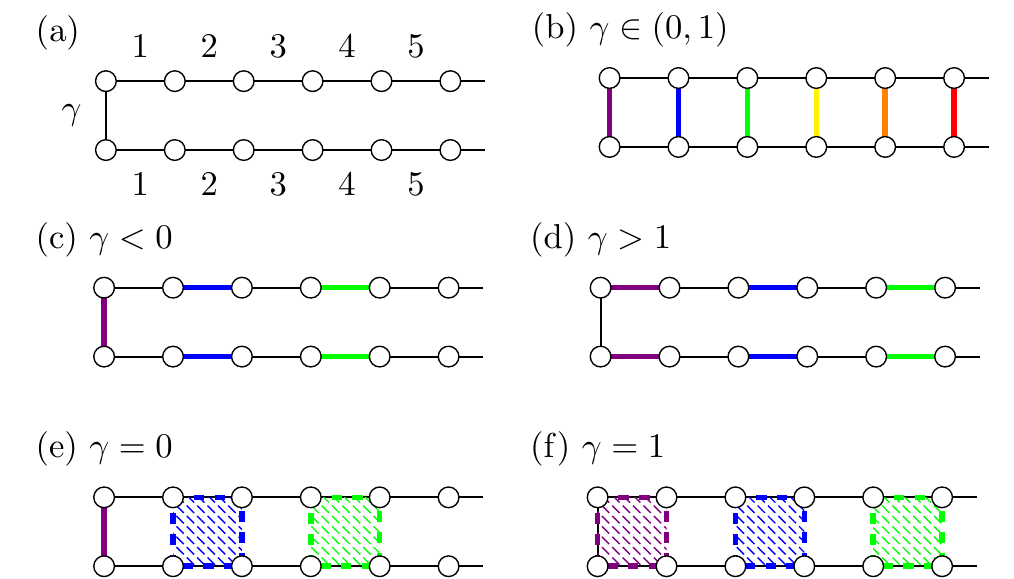}
  \end{center}
  \caption{(a) Folding the rainbow into a two-rung ladder; (b) folded
    rainbow structure, obtained for $\gamma\in(0,1)$; (c) and (d)
    folded dimerized structures, for $\gamma<0$ and $\gamma>1$; (e)
    and (f) folded versions of the transition points, with the plaquettes
    marked where operators $d^\dagger$ act, see Eq. \eqref{eq:bond4c}.}
  \label{fig:folding}
\end{figure}

The structure of the different phases of the rainbow system with a
defect can be properly understood if we {\em fold} the chain around
the central link, as it is shown in Fig. \ref{fig:folding} (a),
converting the chain into a two-rung ladder where sites $+k$ and $-k$
face each other \cite{Samos.19}. This transformation converts rainbow
bonds into vertical bonds and the remaining local bonds into
horizontal bonds. The lower panels of Fig. \ref{fig:folding} present
the bond structure as a function of $\gamma$.

\subsection{Energies}

Let us consider the single-body energy levels $E_k(h,\gamma)$ ($k \in \{0,\cdots,N-1\}$) of $H_L(h,\gamma)$,
, obtained by diagonalizing the corresponding hopping matrix. Due to the
particle-hole symmetry, $E_k=-E_{N-k}$, we need only
consider values up to $L-1$. For large $h$, these single-body energy
levels correspond to the couplings associated with each valence bond,
thus leading us to propose that the following limits are finite,

\begin{equation}
  \lim_{h\to\infty} -{\log|E_k(h,\gamma)|\over h} =
  \Xi_k(\gamma).
  \label{eq:energy_levels}
\end{equation}

\begin{figure}[h!]
\includegraphics[width=7.5cm]{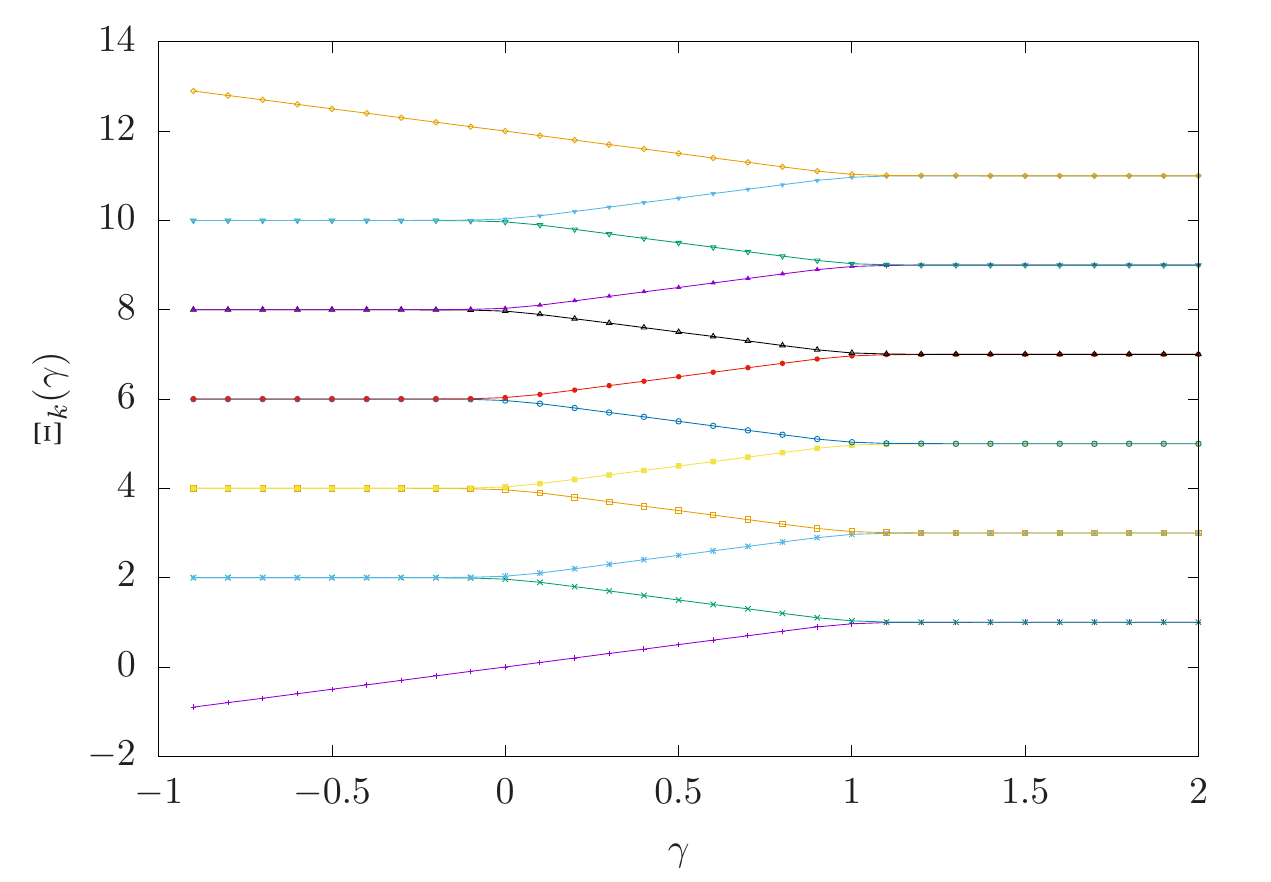}
\includegraphics[width=8cm]{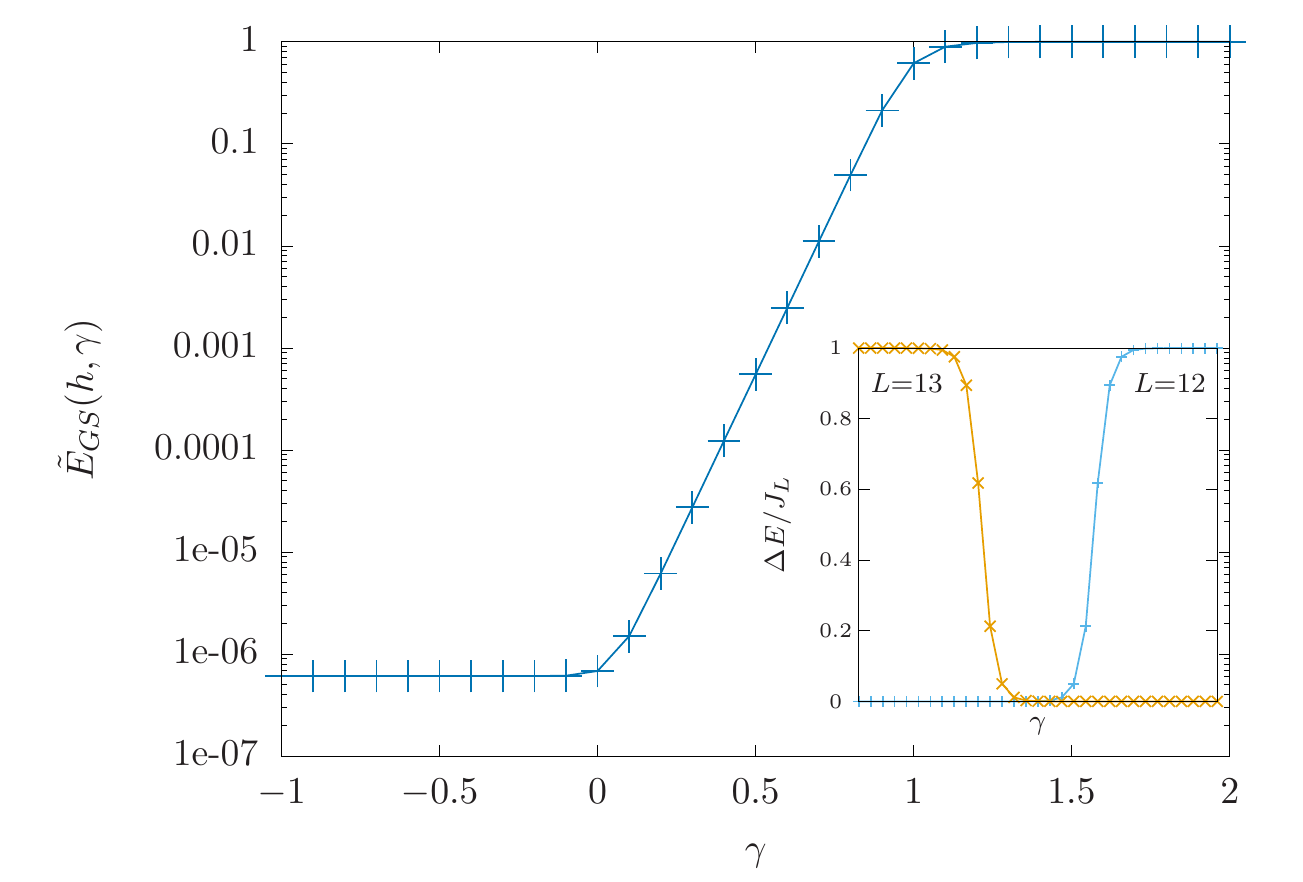}
\caption{Top: plot of $\Xi_k(\gamma)$, obtained numerically for
  $L=12$ and $h=15$, see Eq. \eqref{eq:energy_levels}. Each curve
  matches the renormalized log-couplings along the SDRG
  procedure. Bottom: ground state energy, with the lowest single-body
  energy level removed, as a function of $\gamma$ for the same
  system. Inset: energy gap in units of the lowest energy scale of the
  system, $\Delta E/J_L$, for $L=12$ and $L=13$.}
  \label{fig:energy}
\end{figure}

Fig. \ref{fig:energy} (top) plots these values, $\Xi_k(\gamma)$ as a
function of $\gamma$ for $L=12$, obtained numerically using $h=15$
(for which convergence has been achieved and ). Notice the clear
pattern: for $\gamma>1$, all energy levels are degenerate,
$\Xi_{2k}(\gamma)=\Xi_{2k+1}(\gamma)=2k+1$ for $k\in\{0,\cdots,L/2-1\}$, while for $\gamma<0$ all
energy levels are degenerate and constant, except the first and last which vary exponentially with $\gamma$,
$\Xi_{2k-1}(\gamma)=\Xi_{2k}(\gamma)=2k$ for $k\in\{1,\cdots,L/2-1\}$. Indeed, these values
correspond to the energies associated to the successive valence bonds
of the dimerized phases. On the other hand, for $\gamma\in (0,1)$ the
energy levels are not degenerate, and we can observe the same
alternation of the renormalized log-couplings that we observed in the
SDRG description: $\Xi_k(\gamma)=k+1/2+(-1)^k(\gamma-1/2)$. Thus, the {\em
  transition points}, $\gamma=0$ and $\gamma=1$, correspond to the points
where the degeneracy starts and ends.

The ground state energy is the sum of the energies of the
occupied orbitals, $E_{GS}(h,\gamma)=\sum_{k=0}^{L-1}
E_k(h,\gamma)$. Notice that for large $h$ and $\gamma>1$, the lowest single-body energy $E_0(h,\gamma)$ is the main contribution to $E_{GS}(h,\gamma)$ as its value grows exponentially with $\gamma$ (see the lowest line of the top panel of Fig. \ref{fig:energy}), so we have considered instead the quantity

\begin{equation}
  \tilde{E}_{GS}(h,\gamma)=-E_{GS}(h,\gamma)+E_0(h,\gamma)=-\sum_{k=1}^{L-1} E_k(h,\gamma).
  \label{eq:gsmodified}
\end{equation}
The values of $\tilde{E}_{GS}(h,\gamma)$ are plotted in Fig. \ref{fig:energy} (bottom) for
the same system $L=12$ and $h=15$, in logarithmic scale. Notice the
three regions: for the dimerized phases, $\tilde{E}_{GS}(h,\gamma)$ stays constant, while
for the rainbow phase it grows exponentially. Indeed, for
$h\to\infty$, the energy curve $\log(\tilde{E}_{GS}(h,\gamma))/h$ becomes
non-smooth at $\gamma=0$ and $\gamma=1$, pointing at a phase
transition.

In addition, the inset of Fig. \ref{fig:energy} (bottom) plots the
energy gap $\Delta E/J_L=\left(E_{L}-E_{L-1}\right)/J_L$, normalized with the lowest energy
scale of the system (the lowest coupling constant). We can see that it
presents two types of behaviors, depending wether the spectrum has a long range mode (with $E_{L-1}(h,\gamma)=e^{-Lh}$): for even $L$ it is close to zero ($\Delta E/J_L\sim e^{-h}$) for
$\gamma<1$, while for odd $L$ it is close to zero ($\Delta E/J_L\sim e^{-h}$) for
$\gamma>0$. For $\gamma\in [0,1]$, it is close to zero for all
sizes.

\subsection{Correlations and order parameters}

In order to provide further support to our idea that there is a phase
transition at $\gamma=0$ and $\gamma=1$ in the strong coupling limit,
let us provide two order parameters, that we will call the {\em
  dimerization parameter}, $\Delta_d$ and the {\em rainbow parameter},
$\Delta_r$,

\begin{eqnarray}
  \Delta_d&=&
  \frac{1}{N}\sum_{i=-L+\frac{1}{2}}^{L-\frac{1}{2}}|\ev{c^\dagger_ic_{i+1}}{\psi}|\\
  \Delta_r&=&
  \frac{1}{L}\sum_{i=-\frac{1}{2}}^{L-\frac{1}{2}}|\ev{c_i^\dagger c_{-i}}{\psi}|.
\end{eqnarray}

\begin{figure}
\includegraphics[width=8cm]{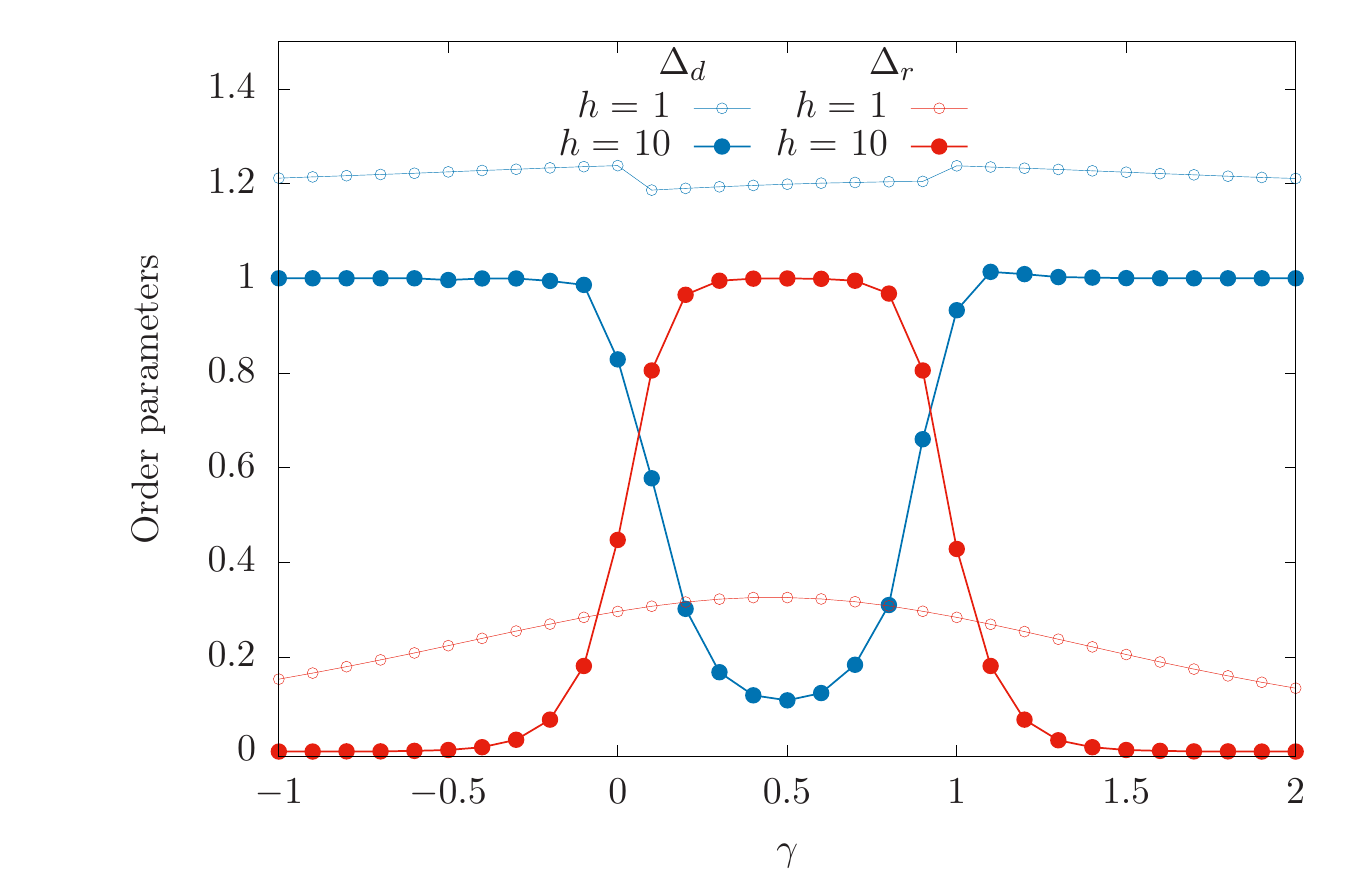}
\caption{Dimerization parameter $\Delta_d$ (blue curves) and rainbow
  parameter $\Delta_r$ (red curves) as a function of $\gamma$, for a
  system of $L=10$. As the inhomogeneity parameter $h$ grows,
  $\Delta_d$ approaches one in the dimerized phases and zero in the
  rainbow phase, while $\Delta_r$ approaches one in the rainbow phase
  and zero in the dimerized phases.}
  \label{fig:orderparam}
\end{figure}

Fig. \ref{fig:orderparam} shows the behavior of these two order
parameters as a function of $\gamma$, for two values of $h$ and
$L=10$. For large $h$ ($h=10$ in the figure), we see that the rainbow
parameter $\Delta_r$ tends to 1 in the rainbow phase
($\gamma\in(0,1)$), while it decays to zero in the dimerized
phases. The opposite behavior is true for the dimerization parameter
$\Delta_d$.

\subsection{Entanglement entropy}

Given a system in a pure state, $\ket{\psi}$, the entanglement entropy
(EE) of a block $A$ is defined as the von Neumann entropy of its
associated reduced density matrix
$\rho_A=\Tr_{\bar{A}}\ket{\psi}\bra{\psi}$, where $\bar{A}$ is the
complementary of $A$.

\begin{equation}
S[\rho_A]=-\Tr \rho_A\log(\rho_A),
\label{eq:von_neumann}
\end{equation}
while the Rényi entropy of order $n$ is defined as

\begin{equation}
S^{(n)}[\rho_A]=\frac{1}{1-n}\log \Tr \rho_A^n.
\label{eq:renyi}
\end{equation}
Needless to say, the different entanglement entropies are determined
by the eigenvalues of the reduced density matrix, also known as {\em
  entanglement spectrum} (ES). There is well known procedure
\cite{Peschel.03} in order to obtain the ES through the spectrum of
the correlation matrix restricted to the block, $\<c^\dagger_i c_j\>$,
with $i$, $j\in A$. The correlation matrices can be exactly obtained
in the strong coupling limit, as it is shown in Appendix
\ref{sec:corrmatrix}. On the other hand, when the state is a VBS, we
can evaluate the EE just by counting the number of bonds which are
broken when we detach the block from its environment, and multiplying
by $\log(2)$, and the same is true for all Rényi entropies.

We have considered two different types of blocks: {\em lateral
  blocks} start from the extreme of the chain, while {\em central
  blocks} are symmetric with respect to the center. In the next
paragraphs we will describe the behavior of their entanglement.

\subsubsection{Lateral blocks. Half chain entropies} 
\label{ssub:boundary_blocks}

Lateral blocks $A_\ell=\{-L+\frac{1}{2},\dots,-L+\frac{1}{2}+\ell\}$
are contiguous blocks containing one of the extremes of the
chain. Concretely, we will be interested in the EE of the half chain,
$S(L)=S[A_L]$ in the strong coupling regime for different values of
$\gamma$. Let us remind the reader that we will only consider even $L$
for simplicity, and that the different phases can be visualized either
in Fig. \ref{fig:strong} and Fig. \ref{fig:folding}, where the blocks
contain $\ell$ sites from the upper leg, starting from the right end.

\begin{itemize}

\item {\em Rainbow phase, $\gamma\in (0,1)$:} the EE (and all other
  Rényi entropies) are merely proportional to the lenght up to
  $\ell=L$, $S[A_\ell]_{\gamma\in(0,1)}=\log(2)\min(\ell,2L+1-\ell)\,$.
  
\item {\em Dimerized phases, $\gamma<0$ or $\gamma>1$:} the lateral
  blocks cut either zero or one bonds for $\gamma>1$,
  $S[A_\ell]_{\gamma>1}=\log(2)(1-(-1)^\ell)/2$; yet, for $\gamma<0$ there is
  always a long-distance bond joining both ends, thus
  $S[A_\ell]_{\gamma<0}=\log(2)\left(1+(1+(-1)^\ell)/2\right)$.
  
\item {\em Transition cases: $\gamma=0$ and $\gamma=1$:} The state is
  not a VBS, so the EE of a block can not be evaluated just by
  counting broken bonds. As we can see in the folded view,
  Fig. \ref{fig:folding}, the sites are grouped into plaquettes
  (except, maybe, for the extremes and the central link). Cutting one
  of these plaquettes horizontally in half contributes a finite amount
  of entanglement $S_a$, which is exactly evaluated in Appendix
  \ref{sec:corrmatrix} (see Eq. \eqref{eq:bond4c}):
  
\begin{equation}
  S_a=\log 20-\frac{4\tanh^{-1}\(\frac{2}{\sqrt{5}}\)}{\sqrt{5}}
  \approx 0.4133,
\label{eq:Sa}
\end{equation}
we are thus led to exact expressions for the half-chain entropy:
$S[A_L]_{\gamma=1}=S_a\,L/2\log(2)$, $S[A_L]_{\gamma=0}=S_a\,(L/2-1)+2\log(2)$.

\end{itemize}

\begin{figure}[h!]
\includegraphics[width=8cm]{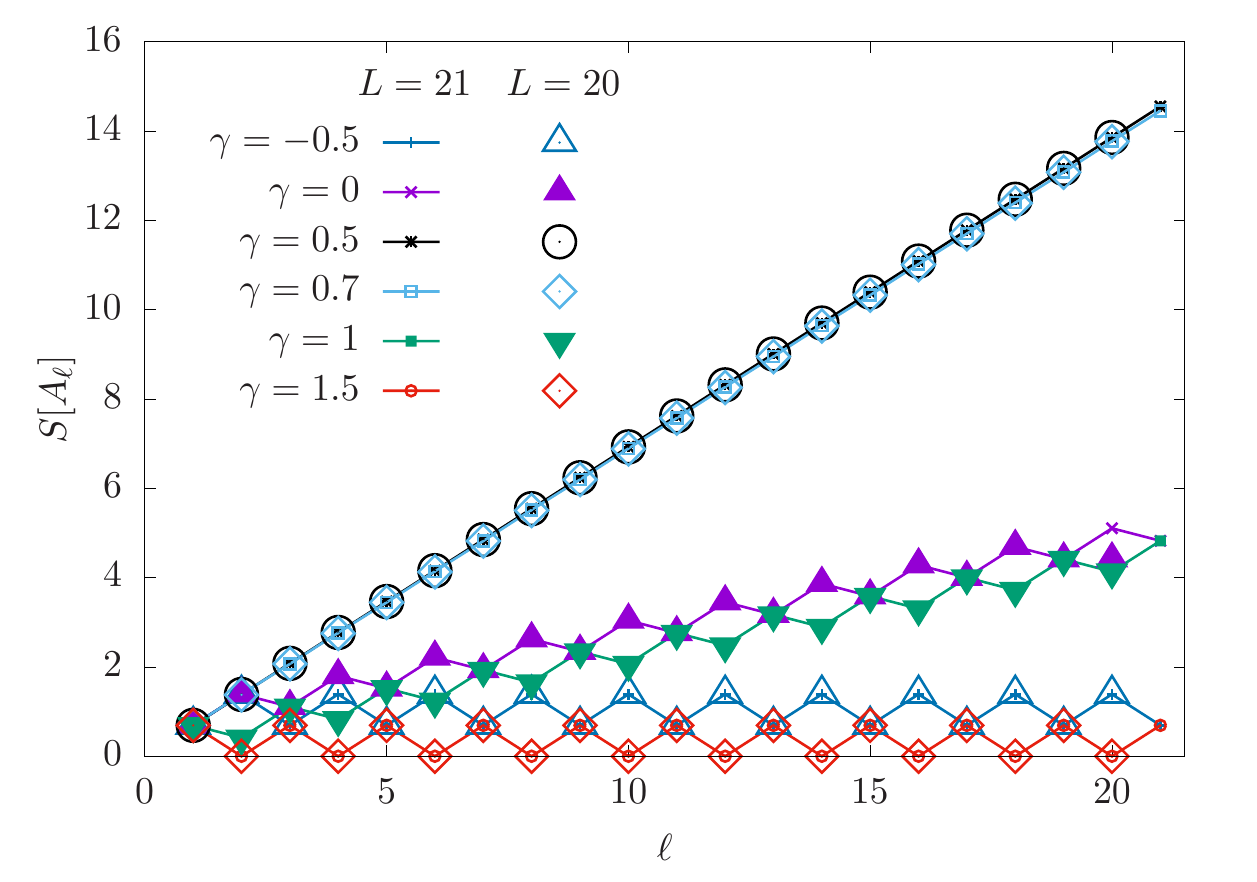}
  \includegraphics[width=8.25cm]{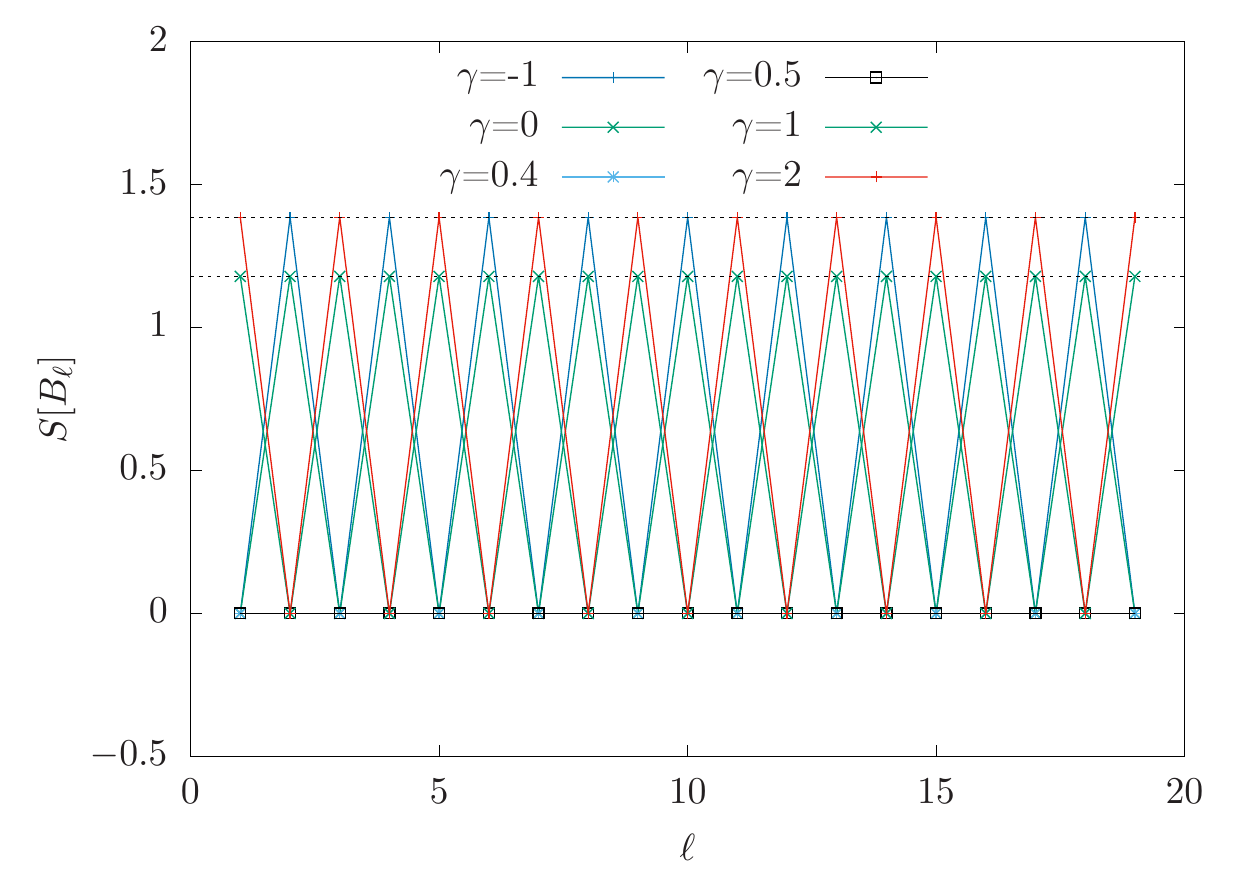}
  \caption{Top: Entanglement entropy of blocks of size $\ell$ using
    $h=10$ for two systems, of size $L=20$ and $L=21$, for different
    values of $\gamma$. Bottom: EE of the central blocks $B_\ell$ for
    $L=20$ sites and $h=10$. The top horizontal line marks $2\log 2$,
    and the lower one marks $S_b$, see
    Eq. \eqref{eq:Sb}}.
\label{fig:s_strong}
\end{figure}

All these results can be checked in Fig. \ref{fig:s_strong} (top) for
two rainbow chains with $L=20$ and $L=21$, using $h=10$, where
$S[A_\ell]$ is plotted as a function of $\ell$ for different values of
$\gamma$. We can see that the $\gamma=-0.5$ and $\gamma=1.5$ cases
show a properly dimerized behavior, and the $\gamma=0.5$ values
correspond to the rainbow, linear with (maximal) slope $\log(2)$. For the
transition points, $\gamma=0$ and $\gamma=1$ we can observe a linear
behavior (with parity oscillations) with a slope $S_a$.

\subsubsection{Central blocks} 
\label{ssub:central_blocks}

In this subsection we consider the EE of central blocks, symmetrically
placed around the center of the chain,
$B_\ell=\{-\ell+\frac{1}{2},\cdots,\ell-\frac{1}{2}\}$. Again, we
suggest the reader to refer to the corresponding panels of
Fig. \ref{fig:strong} and Fig. \ref{fig:folding}, where the blocks now
include $\ell$ rungs starting from the left extreme.

\begin{itemize}

\item {\em Rainbow phase, $\gamma\in (0,1)$:} we always have
  $S[B_\ell]_{\gamma\in(0,1)}=0$.

\item {\em Dimerized phases, $\gamma<0$ or $\gamma>1$:} central blocks
  either cut zero or two bonds. Always usign even $L$ we have
  $S[B_\ell]_{\gamma<0}=(1+(-1)^\ell)\log(2)$ and
  $S[B_\ell]_{\gamma>1}=(1-(-1)^\ell)\log(2)$. 

\item {\em Transition phases, $\gamma=0$ or $\gamma=1$:} central blocks
  can cut plaquettes in half vertically, in the folded view. Each such
  cut contributes a finite amount of entanglement, given by (See
  Appendix \ref{sec:corrmatrix} and Eq. \eqref{eq:bond4c}):

  \begin{equation}
    S_b= \log 5-\frac{\coth^{-1}\(\sqrt{5}\)}{\sqrt{5}}\approx 1.1790,
    \label{eq:Sb}
  \end{equation}
  which leads us to the expressions
  $S[B_\ell]_{\gamma=0}=(1+(-1)^\ell)S_b$ and
  $S[B_\ell]_{\gamma=1}=(1-(-1)^\ell)S_b$.
\end{itemize}

All these features can be checked in Fig. \ref{fig:s_strong} (bottom),
where we can see the central blocks entropy $S[B_\ell]$ as a function
of $\ell$ for different values of $\gamma$. Note that the EE of the
central blocks is always bounded, thus obeying the area law. Non-local
fermionic excitations of the type $b_{i,-i}$ and $d_i$ (see
Eq. \eqref{eq:bonding} and Eq. \eqref{eq:bond4c}) are made local by
the folding operation previously discussed \cite{Samos.19}, and allows
to describe the system state using only {\em short range entanglement}
(SRE).

It is important to realize that it is possible to find local blocks
whose EE is zero independently of the defect (see
Fig. \ref{fig:folding}). This means that the system is topologically
trivial for all $\gamma$ \cite{Samos.19}. On section
\ref{sec:coexistence} we will discuss the site centered symmetry case,
that presents interesting topological features.


\section{Weak inhomogeneity: defect on a deformed background}
\label{sec:weak}

It is relevant to ask whether the phases described in the strong
inhomogeneity limit and the corresponding entanglement transitions
extend into the weak inhomogeneity regime. The answer is no, but some
relevant traits do.

\begin{figure}[h!]
  \includegraphics[width=8cm]{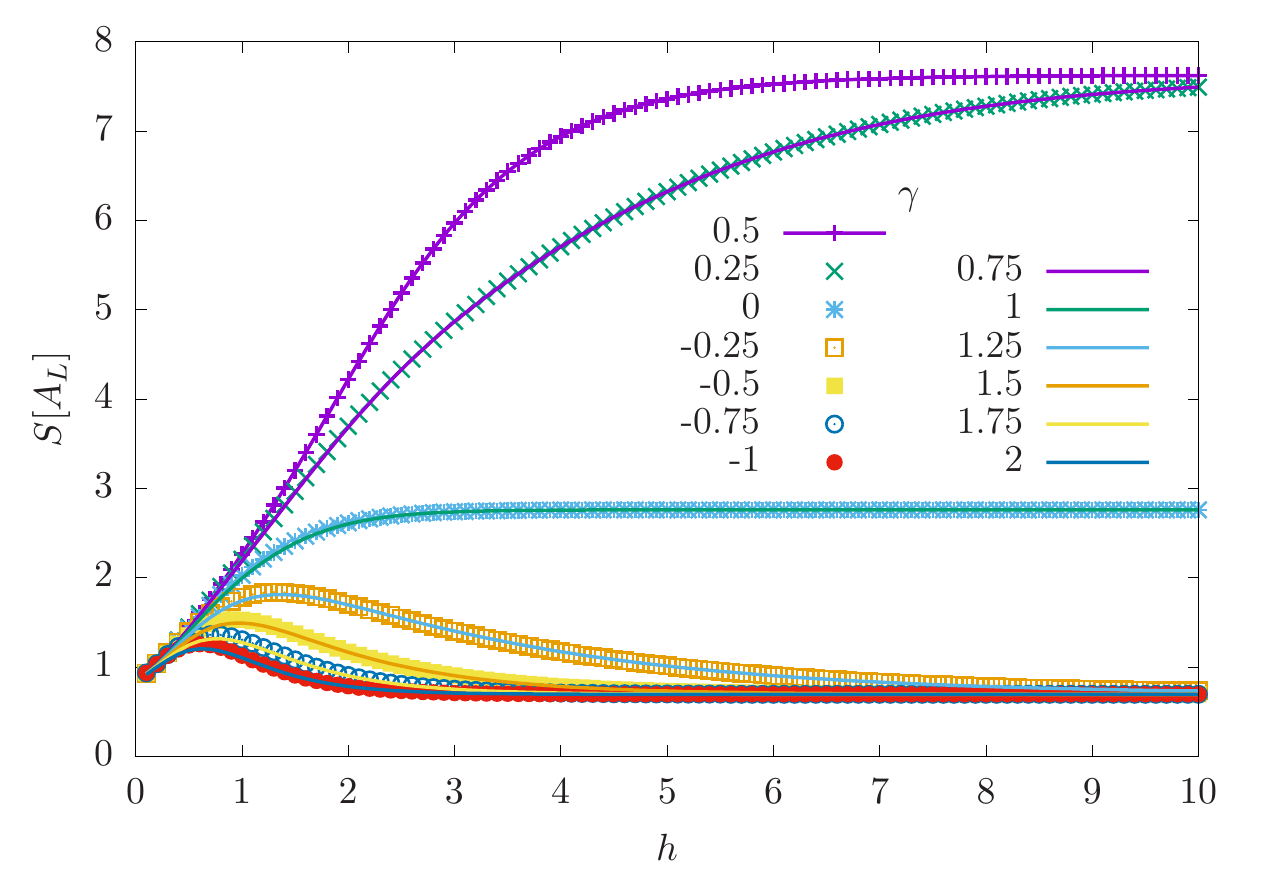}
  \caption{EE of the half system $N=22$ as a function of $h$ for
    different values of the defect strength $\gamma$.}
  \label{fig:runingh}
\end{figure}

In Fig \ref{fig:runingh} we show the dependence on $h$ of the EE of
the half chain, $S(L)=S[A_L]$, for different values of $\gamma$. We
can observe a perfect symmetry between $\gamma$ and $1-\gamma$, and
the three different trends in the large $h$ limit that we have
explained on the previous section: for $\gamma\in (0,1)$ the EE
reaches its maximal value; for $\gamma \in \{0,1\}$, it reaches an
intermediate value ($S_aL/2$); for $\gamma \not\in [0,1]$, it stays at
$\log(2)$. Interestingly, the behaviour is remarkably different for
lower values of $h$, as we will discuss.

For $h=0$ the Hamiltonian \eqref{eq:ham} becomes the standard massless
free-fermionic chain with open boundary conditions (OBC) which can be
described at low energies by a conformal field theory (CFT) with
$c=1$. It is interesting to discuss first such a system in presence of
a defect.

\subsection{Homogeneous chain with defect} 
\label{sub:homogeneous_chain_with_defect}

Let us consider an homogeneous free fermionic chain with OBC and a
defect on its central link, parametrized by a coupling parameter
$\tau$,

\begin{equation}
\label{eq:homoham}
H_\tau= -\frac{\tau}{2} c_{-\frac{1}{2}}^\dagger c_{\frac{1}{2}}
-\frac{1}{2}\sum_{n=1/2}^{L-3/2}
c^\dagger_{n}c_{n+1}+ c^\dagger_{-n}c_{-n-1} +\text{h.c.}
\end{equation}
Let us take the continuum limit of \eqref{eq:homoham} and characterize
its low-energy properties by expanding the local operators $c_n$ into
slow left/right moving components $\psi_{\{L,R\}}$ around the Fermi
points, and introducing a physical coordinate $x=an$, with lattice
constant $a\to 0$, while $L\to \infty$ with $\mathcal{L}=aL$ fixed.

\begin{equation}
  \label{eq:continuum_limit}
  c_m\approx\sqrt{a}\left(e^{ik_Fx}\psi_L(x)+e^{-ik_Fx}\psi_R(x)\right).
\end{equation}
The boundary conditions satisfied by the fields at the edge boundaries
$\psi_{L,R}(\pm\mathcal{L})$ are obtained by imposing
$c_{\pm(L+\frac{1}{2})}=0$:

\begin{equation}
  \psi_L(\mathcal{\pm L})=e^{i\pi(L\pm\frac{1}{2})}\psi_R(\mathcal{\pm L}).
\end{equation}
In order to characterize the effect of the defect $\tau$, we need to
distinguish between the fields on the left side $\psi_{L,R}^I$ and the
right side $\psi_{L,R}^{II}$ of the defect, which are related by a
transfer matrix $\psi^I=T\psi^{II}$ (see Appendix
\ref{sec:relation_with_dirac_equation_with_}):

\begin{equation}
\label{eq:transfermatrix}
\begin{pmatrix}
  \psi_L^I\\
  \psi_R^I
\end{pmatrix}
=\frac{1}{2\tau}\begin{pmatrix}\tau^2+1 & -i(\tau^2-1)\\
  i(\tau^2-1)& \tau^2+1\end{pmatrix}\begin{pmatrix}
    \psi_L^{II}\\
    \psi_R^{II}
  \end{pmatrix}.
\end{equation}
It is important to realize that $T$ {\em only} depends on the defect
and a vicinity of radius $a$ (lattice sites $\pm\frac{1}{2}$ and
$\pm\frac{3}{2}$). Also notice that for $\tau=1$,
$T=\mathbb{I}$. Following \cite{Sierra.14}, we can associate this
transfer matrix to the one associated with a massless Dirac fermion
with a $\delta$ term associated to a mass $m$ and to a chiral mass
$m'$,

\begin{equation}
  \label{eq:transfer_general}
  T_D=\frac{1}{1-r^2-r'^2}\begin{pmatrix}1+r^2+r'^2 & 2(ir+r')\\
    2(-ir+r') & 1+r^2+r'^2\end{pmatrix},
\end{equation}
where $r\propto m$ and $r'\propto m'$ are the reflection coefficients
associated to both terms. If we assume $r'=0$ and compare with
Eq. \eqref{eq:transfermatrix} we find that

\begin{equation}
  r=\frac{1-\tau}{\tau+1}.
\end{equation}
Hence, the field theory associated to the homogeneous system in
presence of a defect Eq. \eqref{eq:homoham} is a massless Dirac
fermion with a $\delta$ potential term that mixes the left and right
moving fermions generating a {\em local} mass placed at the center.

\bigskip

The entanglement properties of this system were studied by Eisler and
Peschel \cite{Eisler.10}. The authors used a conformal mapping to the
isotropic $2D$ Ising model to show that the EE of the half chain
presents a logarithmic behaviour, as predicted by CFT, but with a
coefficient that depends on the strength of the defect which they
called {\em effective central charge}:

\begin{equation}
  S(L)={c_\eff\over 6}\log L + c',
\end{equation}
with

\begin{equation}
c_\eff=\frac{6}{\pi^2}I(s),
\label{eq:ceff}
\end{equation}
and $I(s)$ given by (see Eq. (26) of \cite{Eisler.10}):
\begin{eqnarray*}
  I(s)=-\frac{1}{2}\[\left((1+s)\log(1+s)+(1-s)\log(1-s)\right)\log s\right.\\
  \left.+(1+s)\text{Li}_2(-s)+(1-s)\text{Li}_2(s)\],
\end{eqnarray*}
with $s=\sin(2\arctan\tau)$ and $\text{Li}_2(z)$ is the dilogarithm function \cite{Abramowitz}.

\subsection{Field theory of the rainbow model with a defect}
\label{sub:field_theory_of_the_model}

Let us return to our rainbow model with a defect. In order to build
the field theory describing the low energy physics of Hamiltonian
\eqref{eq:ham} in the weak inhomogeneity regime we need to obtain the
transfer matrix $T_{h,\gamma}$ associated to the defect. Since the
defect is local, we will conjecture that $T_{h,\gamma}$ is determined
by the defect and its closest vicinity (see Appendix
\ref{sec:relation_with_dirac_equation_with_}):

\begin{equation}
  T_{h,\gamma}=\frac{1}{2}e^{h(\gamma-\frac{1}{2})}
  \begin{pmatrix}
    e^{-2h(\gamma-\frac{1}{2})}+1 & -i(e^{-2h(\gamma-\frac{1}{2})}-1)\\
    i(e^{-2h(\gamma-\frac{1}{2})}-1) & e^{-2h(\gamma-\frac{1}{2})}+1 
  \end{pmatrix}.
  \label{eq:transfergamma}
\end{equation}
Note that $T_{h,\gamma}=T$ described in Eq. \eqref{eq:transfermatrix} if we define
 \begin{equation}
  \tau=e^{-h(\gamma-1/2)}.
  \label{eq:defect}
\end{equation} 
Notice that the symmetry $\gamma\rightarrow1-\gamma$ described in the previous section is also present in the transfer matrix: $T_{h,1-\gamma}$ is $T_{h,\gamma}$ with opossite signs in the non diagonal terms and that $\tau=1$ if $h=0$ but also if $\gamma=\frac{1}{2}$. This implies that the defect has no effect in
$H_L(h,\frac{1}{2})$ or, in other terms, we will say that the defect
is {\em absent}. Indeed, evaluating the continuum limit of
Eq. \eqref{eq:continuum_limit} over Eq. \eqref{eq:ham} leads to an
effective Hamiltonian \cite{Ramirez.15,Laguna.17b,Tonni.18}:

\begin{equation}
  H \approx i \int_{-\tilde L}^{\tilde L} d\tilde x \[
\tilde\psi^\dagger_L\partial_{\tilde x} \tilde\psi_L - 
\tilde\psi^\dagger_R\partial_{\tilde x} \tilde\psi_R \],
\label{eq:continuum}
\end{equation}
where $\tilde x$ is given by

\begin{equation}
\tilde x \equiv \textrm{sign}(x) {e^{h|x|}-1\over h},
\label{eq:transf}
\end{equation}
and

\begin{equation}
\tilde\psi_{\{L,R\}}(\tilde x)=\({dx\over d\tilde x}\)^{1/2} \psi_{\{L,R\}}(x).
\label{eq:transf_psi}
\end{equation}
Thus, our field theory is the free Dirac field on a background metric
given by:

\begin{equation}
ds^2=-e^{-2h|x|}dt^2 + dx^2,
\label{eq:metric}
\end{equation}
i.e. a static  metric, defined by a local speed of light or
local Fermi velocity $v_F(x)=e^{-h|x|}$.

The metric \eqref{eq:metric} is Weyl equivalent to the flat metric
with Weyl factor $e^{-h|x|}$, equal to the continuum limit of the
hopping amplitudes Eq. \eqref{eq:hop_rainbow}. Moreover, the metric
has a scalar curvature given by $R(x)=2h\delta(x)-h^2$, i.e. except at
the origin, it is an homogeneous manifold with negative curvature that
can be mapped to the Poincaré metric in the upper half-plane
\cite{Laguna.17} or the anti-de Sitter (AdS) metric in 1+1D
\cite{MacCormack.18}. As a consequence, the field theory associated to
the Hamiltonian $H_L(h,\gamma\neq\frac{1}{2})$ for low energies should
be described by a free Dirac theory with a local defect --which is
analogous to the one studied in the previous subsection-- but in the
background metric described above. In what follows we show that this
is the case by studying the entanglement properties such as the
entanglement entropy, the entanglement spectrum, the entanglement
Hamiltonian and the entanglement contour.

\subsection{Entanglement entropy}
\label{sub:entanglement_entropy}

In the case of absence of defect, $\gamma=\frac{1}{2}$, the EE can be
evaluated for intervals of the form $(-L,x)$ within a 2D CFT
\cite{Holzhey.94,Vidal.03,Calabrese.04,Calabrese.09}, leading to the
expression

\begin{equation}
S^{(n)}_\CFT(x)= c {n+1\over 12n} \log Y(x),
\label{eq:scft}
\end{equation}
where

\begin{equation}
Y(x)= {2L\over\pi\epsilon} \sin\({\pi (L+x)\over 2L}\),
\label{eq:Y}
\end{equation}
and $\epsilon$ is the UV cutoff. However, the actual entropy of the
discrete state is not {\em exactly} equal to that because a
non-universal term must be added. Its value is exactly known in the
case of the free-fermionic field \cite{Jin.04,Fagotti.11} and will not
be considered here. We can compute the universal part of the EE by
making an appropriate use of transformation \eqref{eq:transf} on
expression \eqref{eq:scft}. Indeed, besides the transformation of $L$
and $x$, we need to take into account the transformation of the UV
cutoff, $\epsilon$, through the Weyl factor, $\tilde\epsilon= e^{h|x|}
\,\epsilon$ in our metric. We obtain

\begin{equation}
S^{(n)}_{\gamma=\frac{1}{2}}(x)=c{n+1\over 12n} \log \tilde Y(x),
\label{eq:srb}
\end{equation}
with

\begin{equation}
\tilde Y(x) =  e^{-h|x|}
{e^{hL}-1\over \pi h} \cos\( {\pi\over 2}{e^{h|x|}-1\over e^{hL}-1} \).
\label{eq:Ytilde}
\end{equation}
The half chain EE scales linearly:

\begin{equation}
  S_{\gamma=\frac{1}{2}}(L)\approx \frac{c\, hL}{6}.
  \label{eq:hcent}
\end{equation}

\begin{figure}
  \includegraphics[width=8cm]{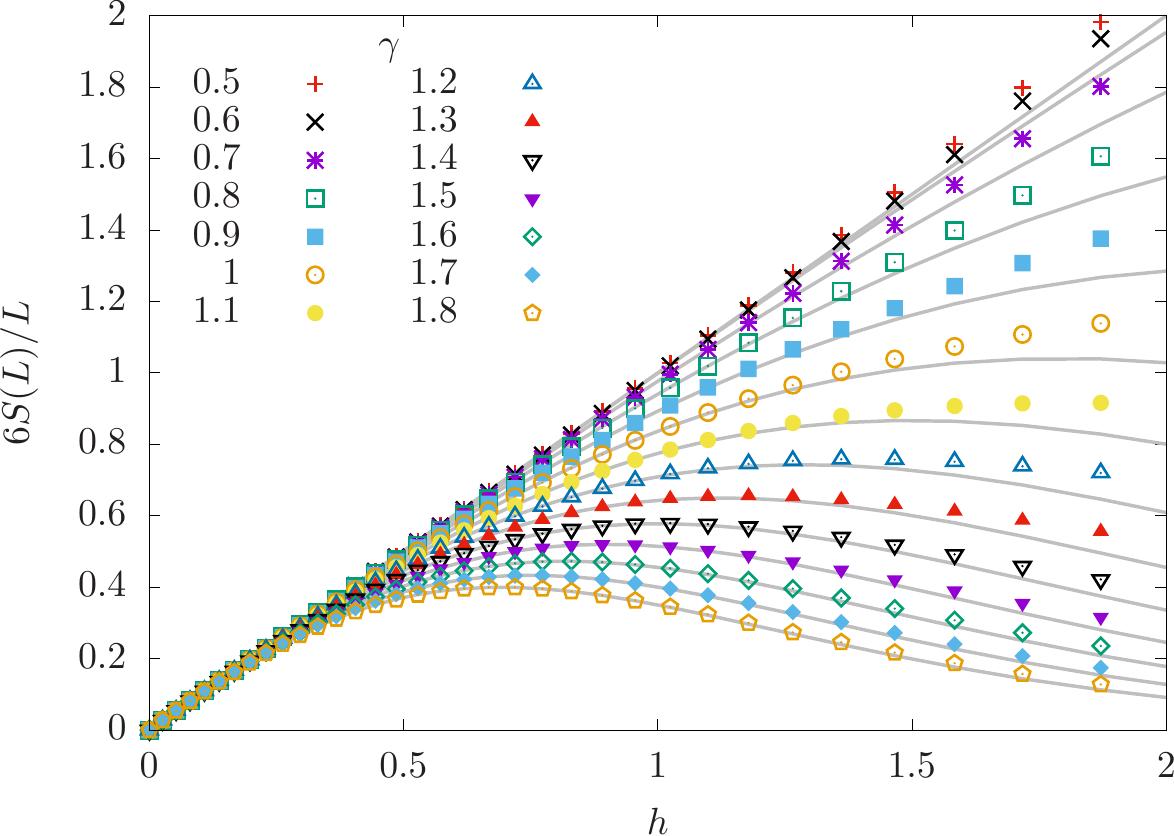}
  \includegraphics[width=8cm]{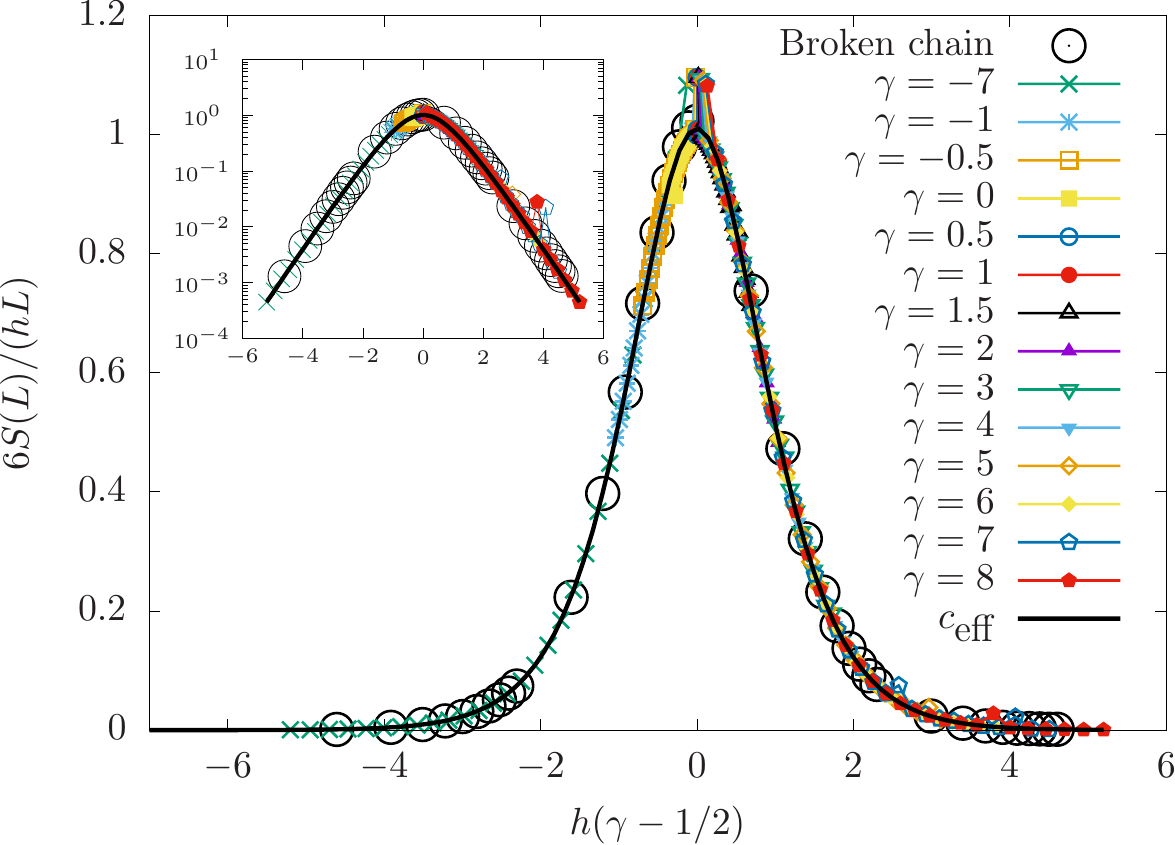}
  \caption{Top: Entropy per site of the rainbow model with a defect,
    $s(h,\gamma)$ as a function of $h$, for different values of
    $\gamma$. Soft continuous lines correspond to the theoretical
    prediction, Eq. \eqref{eq:hc2}. Bottom: entropy per site divided
    by the inhomogeneity parameter, as a function of the defect intensity,
    $h(\gamma-1/2)$, showing the collapse predicted by
    Eq. \eqref{eq:hc2}.}
  \label{fig:peschel}
\end{figure}

However, the defect ($\gamma\neq\frac{1}{2}$) creates a mass and
introduces a scale, breaking the conformal invariance of the
system. As a consequence, the previous formulae can not be applied to
compute the EE. Nevertheless, the EE should follow
Eq. \eqref{eq:ceff}, with the modifications associated to the change
of background. Indeed, we should modify Eq. \eqref{eq:hcent} as
\begin{equation}
  S_{\gamma}(L)\approx \frac{c_\eff(\tau)\, hL}{6}
  \label{eq:hc2}
\end{equation}
where $\tau$ is given by Eq. \eqref{eq:defect}. In order to check this,
we have obtained the {\em entropy per site}, defined for convenience
as
\begin{equation}
  s(h,\gamma)=\lim_{L \to\infty} {6S[A_L]\over L}.
\end{equation}
The values of $s(h,\gamma)$ are obtained through a linear fit. Fig. \ref{fig:peschel} (top) shows this entropy per
site as a function of $h$ for several values of $\gamma$. For very low
values of $h$, all curves seem to collapse. Yet, for
$\gamma\not\in[0,1]$, the curve $s(h)$ eventually presents a maximum
and decays to zero. This is a signature that the system will obey the
area law in the strong inhomogeneity limit. The validity of
Eq. \eqref{eq:hc2} can be checked with the soft continuous lines,
which correspond to the theoretical prediction. Indeed, for low values
of $h$ the prediction is very accurate, losing this accuracy for
large inhomogeneity ($h\approx 1.5$).

Furthermore, Eq. \eqref{eq:hc2} suggests that the entropy per site will
collapse if we plot $s(h,\gamma)/h$ as a function of a measure of the
defect intensity, $h(\gamma-1/2)$. Indeed, this collapse can be
seen in the bottom panel of Fig. \ref{fig:peschel}, showing the
universal curve for $c_\eff(\tau)$. The high accuracy of this collapse
can be checked in the inset, which shows the same data in logarithmic
scale. Moreover, the circles correspond to the plot of $c_\eff$ in
Eq. \eqref{eq:ceff} as a function of $\log(\tau)$, for comparison.

\subsection{Phase Diagram} 
\label{sub:phase_diagram}

\begin{figure}
  \includegraphics[width=8cm]{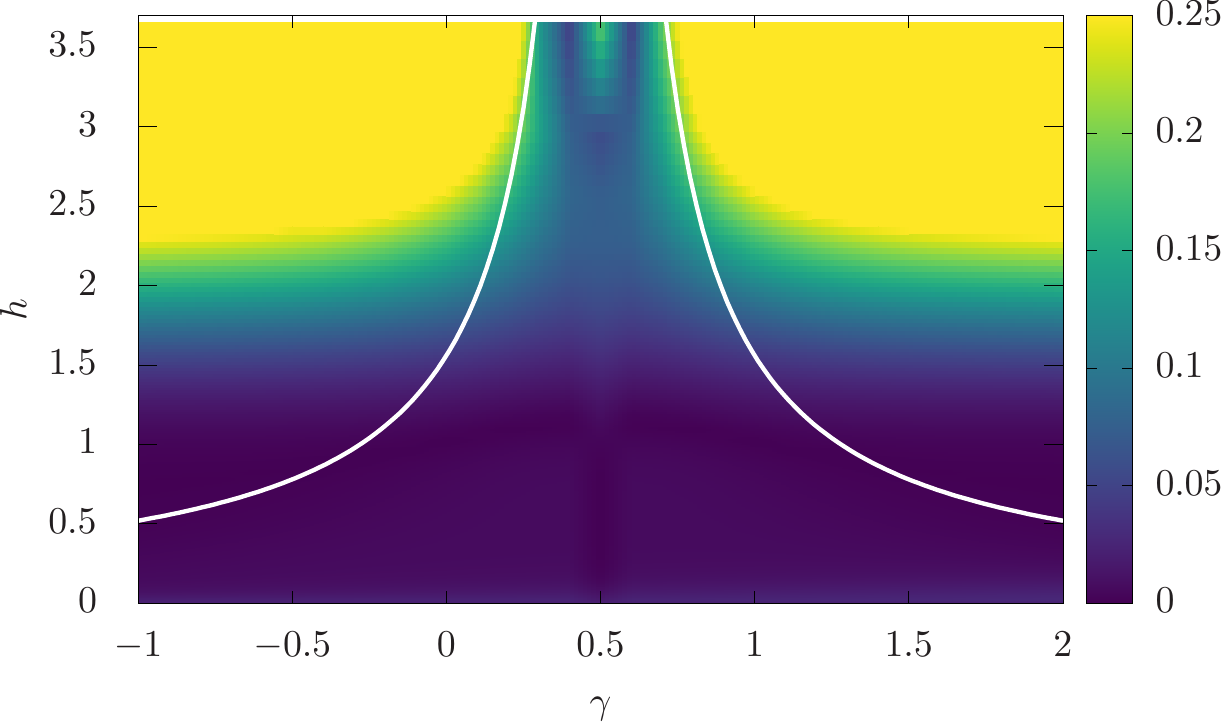}
  \caption{Phase diagram, absolute error of the EE prediction;
    the white lines correspond to the local maximum of the entropy
    density, $s(h)$.}
  \label{fig:error}
\end{figure}

In Fig. (\ref{fig:error}) we show the relative error between the theoretical prediction and the numerics 
\begin{equation}
	\delta s(h,\gamma)=\frac{|s(h,\gamma)-hc_{\text{eff}}|}{s(h,\gamma)},
\end{equation}
 in the color intensity. The
white lines correspond to the theoretical values of the relative
maxima of $s(h,\gamma)$ as a function of $h$, following
Eq. \eqref{eq:hc2}. Notice that the theoretical prediction states
that, for all $\gamma$, the curve $s(h)$ will present a maximum and
decay to zero afterwards. Thus, weak inhomogeneity regime presents a smooth crossover into the three phases of the strong coupling regime described in the previous section. or large $h$ lattice effects become dominant and the universal properties predicted by the field theory approach are lost.

\subsection{Beyond Entanglement Entropy} 
\label{sub:beyond_entanglement_entropy}

The characterization of entanglement can be improved with the study of
the entanglement spectrum (ES), entanglement contour and entanglement
Hamiltonian. All these mathematical objects are associated to the
reduced density matrix of the block, $\rho_A$.

\subsubsection{Entanglement Hamiltonian} 
\label{ssub:entanglement_hamiltonian}

The reduced density matrix $\rho_A$ of a Gaussian fermionic state can
be written in the form

\begin{equation}
\rho_A = \exp(-2\pi H_A) \equiv \exp\(-2\pi\sum_p \epsilon_p d^\dagger_p d_p \),
\label{eq:rhoA}
\end{equation}
for some fermionic operators $d_p$. The $\epsilon_p$ are called the
single-body entanglement spectrum, but the term {\em single-body} is
usually dropped. Operator $H_A$ is termed the {\em entanglement
  Hamiltonian} (EH), and it can be shown to be approximately local for a
1+1D CFT \cite{Tonni.18}. Indeed, it can be written as

\begin{equation}
H_A \approx \sum_i \beta_A(i)\; c^\dagger_i c_{i+1},
\label{eq:EH}
\end{equation}
where the $\beta_A(i)$ constitute {\em entanglement couplings}, and
can be accounted for in CFT providing an extension of the
Bisognano-Wichmann theorem \cite{Cardy.16,Tonni.18}. There are also
non-zero terms presenting long-range interactions, but they are
expected to be very small. The estimation of the set of $\beta_i$ is
obtained by minimizing an error function $E(\beta)\equiv\sum_{i,j\in
  A}\(C_{ij}-Tr(\rho_A(\beta)c^\dagger_ic_j)\)$ using standard
optimization techniques \cite{Tonni.18}.

\begin{figure}
  \includegraphics[width=8cm]{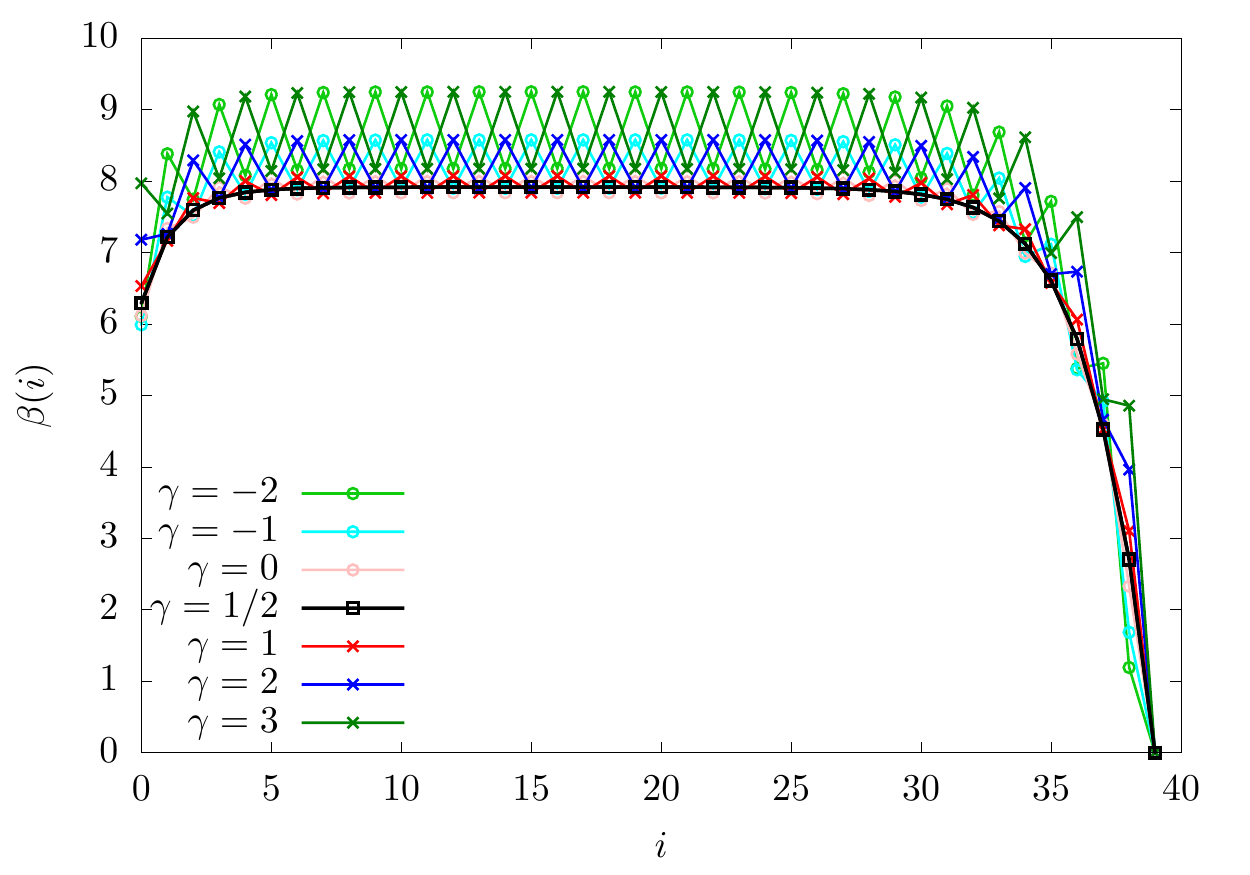}
  \caption{Entanglement Hamiltonian coefficients $\{\beta(i)\}$ for
    the left half of a $L=20$ rainbow with a defect, using
    $h=0.5$. Notice that, for $\gamma=1/2$ the bulk is flat, as
    expected, but for the other values the Hamiltonian coefficients
    present dimerization, which changes the high and low values when
    we change $\gamma$ for $1-\gamma$. }
  \label{fig:eh}
\end{figure}

The numerical values of $\{\beta(i)\}$ for the left half (block $A_L$)
of a $L=20$ system, using $h=0.5$ and different values of $\gamma$ are
shown in Fig. \ref{fig:eh}. For $\gamma=1/2$ the EH of the rainbow
system presents flat coefficients $\beta(i)$ everywhere except near
the physical boundary (left extreme) and near the internal boundary
(right extreme), where it follows the Bisognano-Wichmann prediction,
that they will decay to zero linearly, with slope $1$. Yet, in
presence of a defect we observe an increasing {\em dimerization} of
the EH.

Let us remind the reader that the flat profile for $\{\beta(i)\}$ in
the rainbow case accounts for the fact that the rainbow GS for all
values of $h$ resembles a {\em thermofield double} \cite{Tonni.18},
i.e.

\begin{equation}
  \ket{\Psi} \approx \sum_n \exp(-\beta E_n/2) \ket{n}_L\otimes\ket{n}_R,
  \label{eq:tfd}
\end{equation}
where $E_n$ and $\ket{n}_{\{L,R\}}$ are the energies and eigenstates
of the homogeneous free Hamiltonian on the left/right with open
boundaries. Thus, we are led to the following claim: in presence of a
defect, the ground state of Hamiltonian \eqref{eq:ham} is still
approximately a thermofield double, but of a {\em dimerized}
Hamiltonian, with dimerization parameter associated to the defect
strength $\gamma$.

We would like to stress that the cases of $\gamma$ and $1-\gamma$ are
extremely similar, only interchanging the higher and lower values of
the dimerization pattern.

\subsubsection{Entanglement Spectrum}

For free systems, the entanglement spectrum (ES)
$\{\epsilon_k\}_{k=1}^L$ of a block $A$ is the single-body spectrum of
the reduced density matrix $\rho_A$ \cite{Li_Haldane.08}. In terms of
the ES, the eigenvalues $\{\nu_k\}_{k=1}^L$ of the block correlator
matrix are written \cite{Peschel.03} as:

\begin{equation}
  \nu_k=\frac{1}{e^{\epsilon_k}+1}.
  \label{eq:epsilonk}
\end{equation}
The ES contains more physical information than the entanglement
entropy. In some cases, its low part can be regarded as the energy
spectrum of a boundary CFT \cite{Lauchli2013}.

\begin{figure}
\includegraphics[width=8cm]{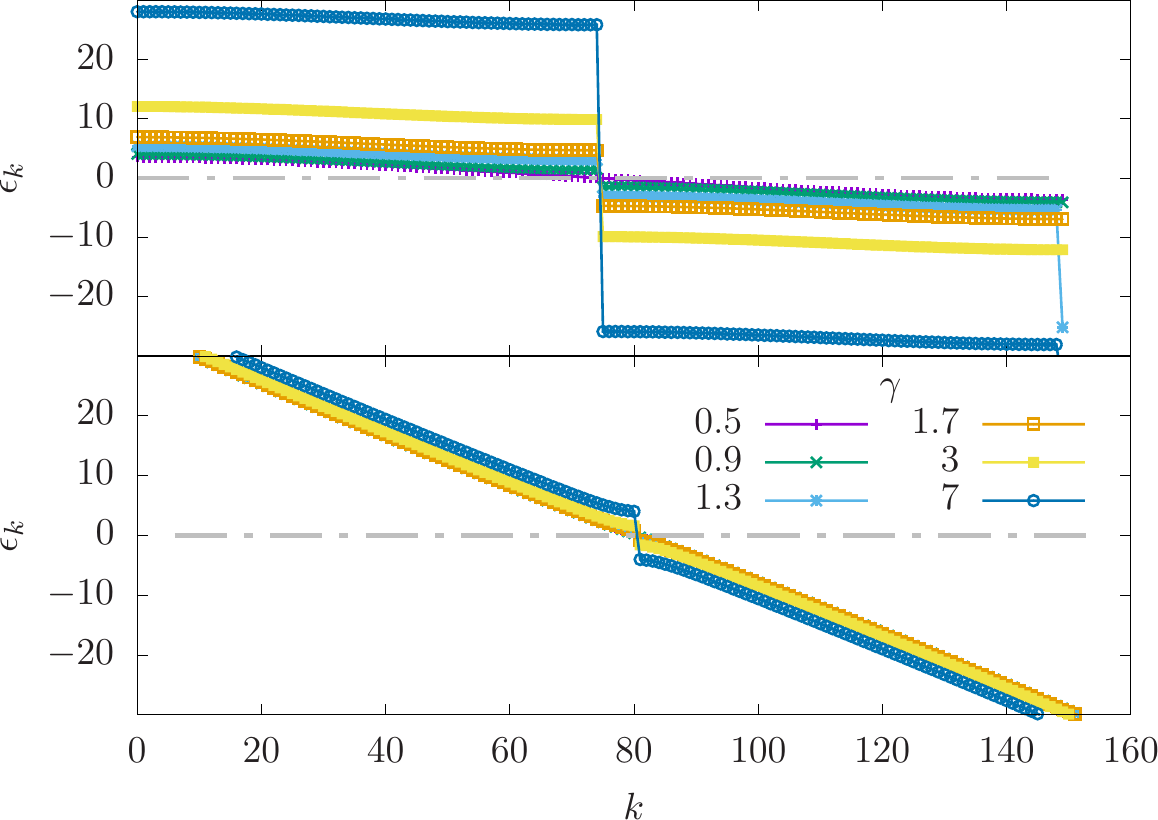}
\includegraphics[width=8cm]{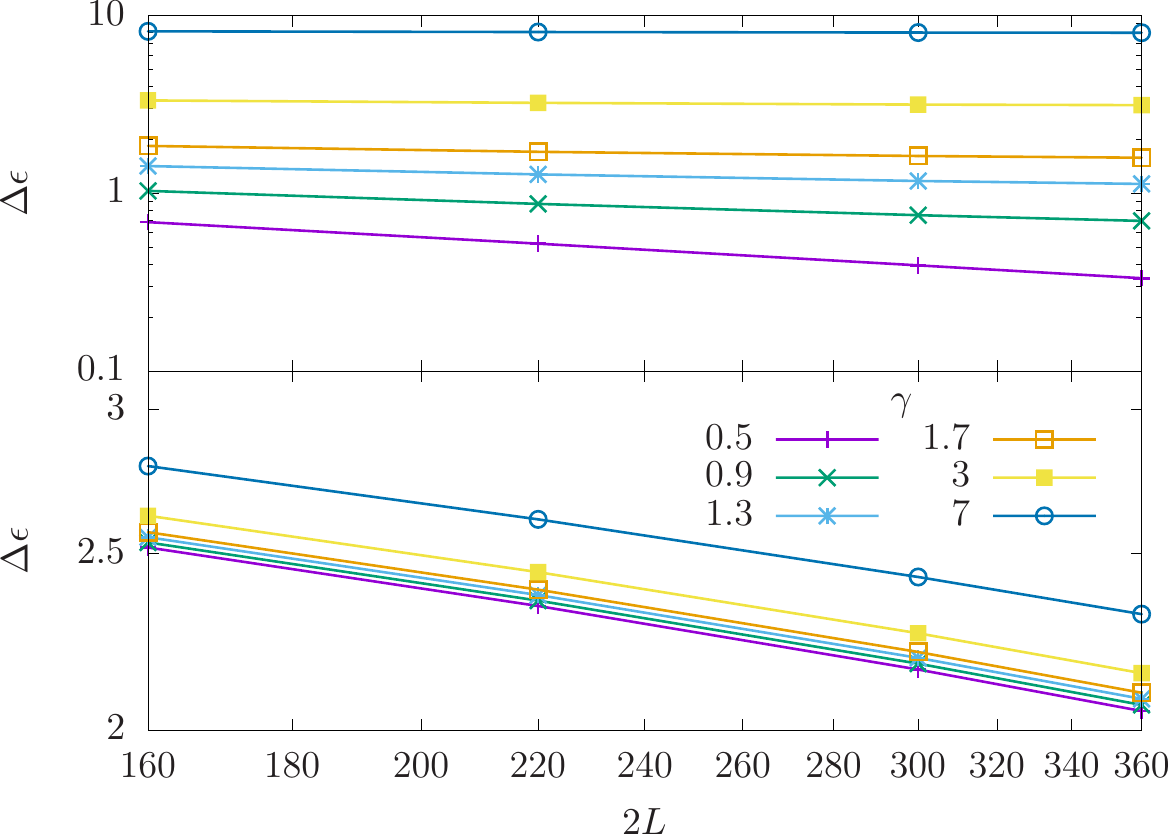}
\caption{Top: Entanglement spectrum $\epsilon_k$ of the left block of
  a system with $L=150$ and different values of $\gamma$. Upper panel:
  $h=0.32$, lower panel: $h=0.015$. Bottom: Scaling of the ES gap
  $\Delta\epsilon$ with size $2L$ for different values of $\gamma$. The
  upper panel shows the case $h=0.32$, and the lower panel $h=0.015$.}
   \label{fig:es}
\end{figure}

We have considered the full ES of the left half block, $A_L$, for
different values of $\gamma$. As it can be expected, the defect
preserves the particle-hole symmetry. The most salient feature is that
the ES shows a finite gap $\Delta \epsilon$ whose width grows with
$\gamma$, as can be seen in Fig. \ref{fig:es} (top). For a CFT
system, the entanglement gap, $\Delta\epsilon\sim 1/\log(L)$, but for
a deformed system such as the rainbow we should consider instead
$\Delta\epsilon \sim 1/\log(\tilde L)\sim 1/L$. Indeed, for low $h$
the gap decays linearly with the system size, as we can see on the
bottom panel of Fig. \ref{fig:es} for $h=0.015$, but it seems to reach
a finite value for $h=0.32$.

\subsubsection{Entanglement Contour} 
\label{ssub:entanglement_contour}

The {\em entanglement contour} \cite{Vidal.14} attempts to answer the
question about where is the entanglement entropy located. The
entanglement entropy of a block is decomposed

\begin{equation}
S_A=\sum_{i\in A} \sigma_A(i),
\label{eq:contour}
\end{equation}
with $\sigma_A(i)\geq 0$. Although the entanglement contour is not
uniquely defined, different candidate definitions have provided very
similar values \cite{Coser.17,Tonni.18,Roy.19}, thus pointing at the
existence of a deeper contour which would have the current candidates
as approximations. Since the rainbow system is defined here for a
free-fermionic system, we will employ the approach given in \cite{Vidal.14}.

\begin{figure}
  \includegraphics[width=8cm]{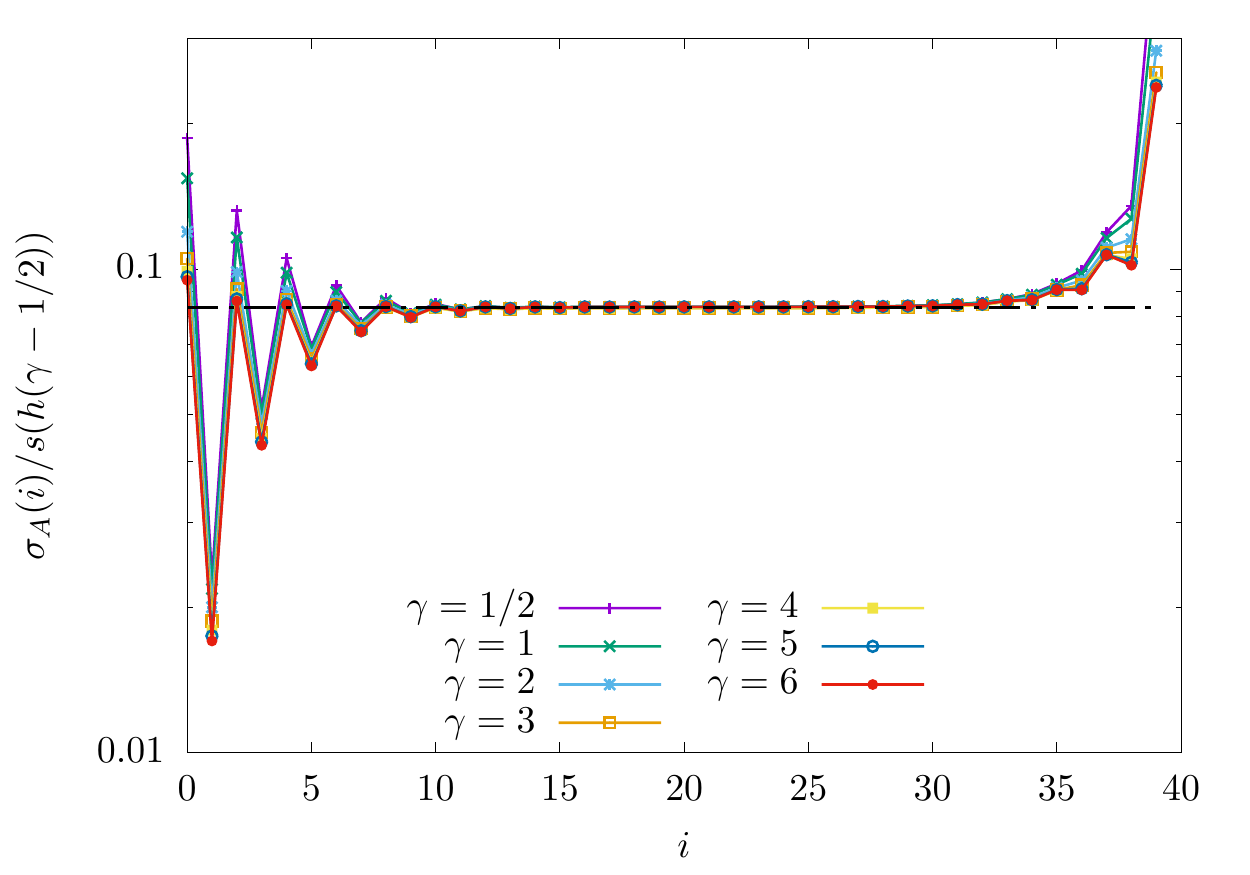}
  \caption{Entanglement contour of the left half (the left edge is the physical boundary while the right one is originated by the block) of the rainbow model
    with a defect using $L=40$ and $h=0.5$, scaled with the entropy
    per site predicted in Eq. \eqref{eq:hc2}, for different values of
    $\gamma$.}
  \label{fig:contour}
\end{figure}

Fig. \ref{fig:contour} shows the curve of the entanglement contour for
the left block of the rainbow GS using $L=40$ and $h=0.5$, for
different values of $\gamma$, scaled with the entropy density
predicted in Eq. \eqref{eq:hc2}. The collapse is very clear in the
bulk region, which presents universal features, and a little bit
less near the boundary, where it does not. Importantly, notice that
the entanglement contour does not present any oscillations related to
dimerization, with a constant entropy per site in the bulk.

\section{Defect in a symmetry protected topological phase}
\label{sec:coexistence}

The system considered so far, Eq. \eqref{eq:ham}, presents {\em bond
  centered symmetry}, i.e.: the center of symmetry is in the middle
point of the central link. However, many different properties arise when
we consider {\em site centered symmetric} (scs) systems
\cite{Samos.19}, where the center of symmetry corresponds to a
site. Let us consider a system defined on a chain with $N=2L$ sites,
whose Hamiltonian is:

\begin{equation}
  H_N(h,\gamma)_\scs=-\frac{1}{2}\sum_{m=1}^N
  J_m c_m^\dagger c_{m+1} + \text{h.c.},
  \label{eq:hamscs}
\end{equation}
where the fermions are now placed on integer positions and there are
two equal central hoppings depending on $\gamma$:

\begin{equation}
J_m = \begin{cases}
  e^{-h(\left|m - (L + \frac{1}{2})\right| - \frac{1}{2})} & \text{if
  } m\neq L,L+1,\\
  e^{ -h\gamma }  & \text{if } m\in\{L,L+1\},
\end{cases}
\label{eq:hop_scs}
\end{equation}
i.e., the log-couplings (see Eq. \eqref{eq:logcoupling}) present the
pattern $\{\dots,3,2,1,\gamma,\gamma,1,2,3,\dots\}$. The site-centered
symmetry manifests itself through the invariance of the hoppings under
an inversion around the central site $L+1$: $J_n=J_{L+1-n}$. Notice
that the new notation is different from the bond centered symmetry
case, Eq. \eqref{eq:ham}, which is now convenient due to the different
type of symmetry.

In \cite{Samos.19} it was shown that after performing a folding
operation around the central site, the system becomes an inhomogeneous
realization of the SSH model, thus belonging to the BDI class of
topological phases \cite{Kennedy1992,Altland1997,P10,Fidkowski2011,Turner2011,W11a,W11b,P12}. Since the topological nature of
the state is highlighted after removing the local
entanglement \cite{W11b}, it is better to study the
system in the strong inhomogeneity regime. The fermionic excitations
are not spread along the whole system as it is the case in the weak
inhomogeneity limit. Hence, we will study the system in the strong
coupling regime $H_N(h\gg1,\gamma)_\scs$ by means of renormalization
schemes that depend on the value of $\gamma$ (see details in Appendix \ref{sec:detailsscs}).

\bigskip

Let us start by considering the case $\gamma\leq1$. The dominant interaction
involves the three central sites, $L$, $L+1$ and $L+2$. With a real space first order perturbation theory RG \cite{Samos.19}. On each step, three fermions are truncated into one which participates on the next step (unlike the RG of the systems with bcs symmetry, where the fermions are integrated out on each step and hence are decoupled from the system) leading to a topological ground state with non removable entanglement that belongs to the BDI class \cite{Altland1997,Samos.19}. The case with $\gamma=1$ differs only on the first step of the RG where 5 spins (instead of three) are truncated to one.\\ 
 On the other hand, the case $\gamma>1$ is again different. Starting from $H_N(h,\gamma)_\scs$, the dominant interactions are two non consecutive log-couplings 1 which allows the use of the Dasgupta-Ma RG Eq. \eqref{eq:sdrg}, leading to an effective system whose Hamiltonian is $H_{N-4}(h,1+\gamma)_\scs$. If $1+\gamma$ happens to be the dominant interaction, three fermions are involved so the Dasgupta-Ma Rg is not applicable anymore and the way of procedure is described in the previous paragraph. On the contrary, if the log-couplings 2 are the dominant interaction, the Dasgupta-Ma RG can be applied again leading to a new Hamiltonian $H_{N-8}(h,2+\gamma)_\scs$. Hence, the same dichotomy is present in the next step. The procedure iterates and unless $\gamma>L-1$ the RG flows eventually to a dominant interaction which involves three fermions.

\bigskip

\begin{figure}
  \includegraphics[width=8cm]{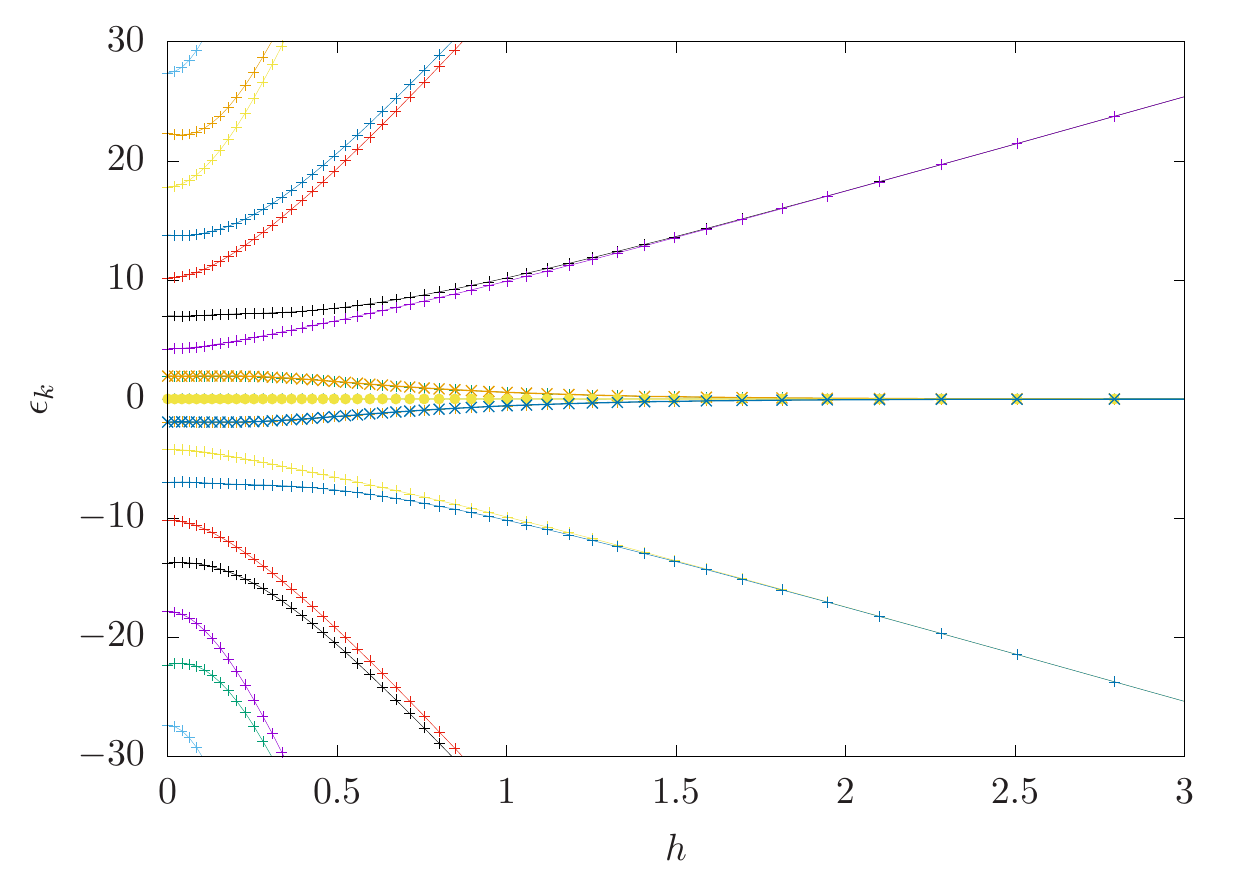}
  \caption{Single body entanglement spectrum $\epsilon_k$ of a block
    $B_{11}$ of a system described by the Hamiltonian
    $H_{46}(h,\frac{17}{2})_\scs$ as a function of $h$. Notice the
    topological zero mode, which does not depend on $h$.}
  \label{fig:fig:cent_sp}
\end{figure}

  Therefore we see that the GS of the Hamiltonian $H_N(h,\gamma>1)_\scs$ is obtained via the application of two kinds of renormalization group schemes. As a consequence, the ground state of this Hamiltonian has two different phases that coexist: a dimerized phase around the defect and the BDI phase
away from it. This coexistence is well captured by considering the EE
of central blocks $B_{\ell}=\{L-\ell,L+2+\ell\}$, with
$\ell\in\{0,\dots, (L-1)\}$. Since the system is topologically
non-trivial, there is entanglement that cannot be removed (the EE in
bounded by $\log2$ for all $B_{\ell}$) and for blocks $B_\ell$ with
$\ell<\floor{\gamma}$, $S{B_\ell}=3\log{2}$ due to the fact that there
are two fermionic excitations
$b_{L-2(\ell-1),L-2(\ell-1)-1}^\dagger\ket{0}$ and
$b_{L+2\ell,L+2\ell+1}^\dagger\ket{0}$ that are not fully contained in
the block $B_\ell$. Furthermore, the dimerized phase appears only in
the strong coupling limit $h\to\infty$ while the BDI phase is
independent of the inhomogeneity parameter. This fact can be checked
by considering the behaviour of the single body entanglement spectrum
$\epsilon_k$, see Fig. \ref{fig:fig:cent_sp} and
Eq. \eqref{eq:epsilonk}. There is a zero mode $\epsilon_0=0$ for all
$h$ that gives rise to a double degeneracy of the many-body
entanglement spectrum and it is a signature of the topological nature
of the state. There are also two additional zero modes due to the
presence of the defect but they are not topological, since they depend
on the inhomogeneity.

\begin{figure}[h!]
\includegraphics[width=8cm]{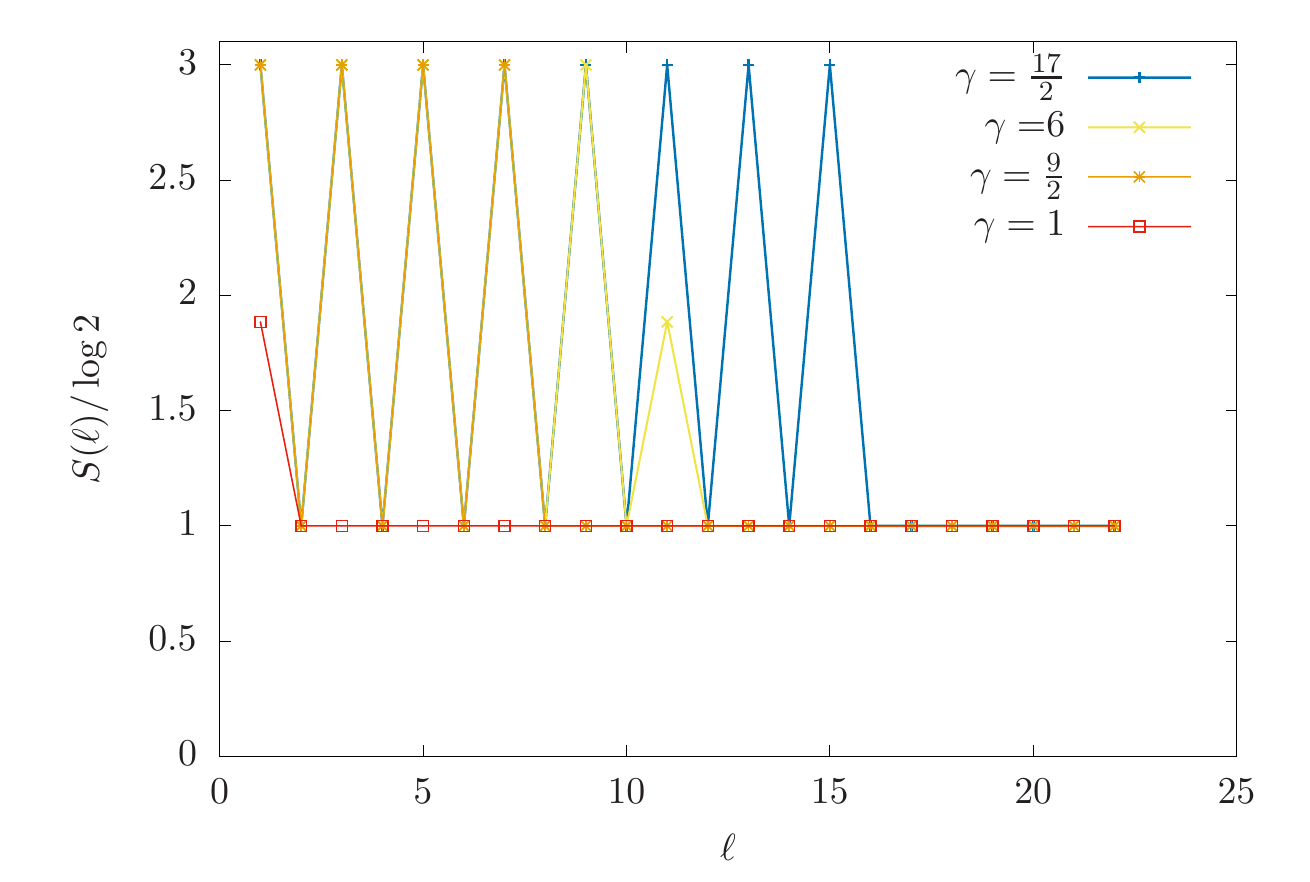}
  \caption{Entanglement entropy of the system with Hamiltonian
    $H_{46}(15,\gamma)_{\text{scs}}$ partitioned with central blocks
    $B_\ell$ for different values of $\gamma$. The entropy is
    $S[B_\ell]=3\log2$ if $\ell<\floor{\gamma}$ and it is
    $S[B_\ell]=\log2$ if $\ell>\floor{\gamma}$. If
    $\gamma\in\mathbb{N}$, $S[B_{\ell=\gamma}]$ takes another value
    which is a consequence of the quadruple tie that we have discussed
    in the text.}
\label{fig:cent_scs}
\end{figure}


\section{Conclusions and further work}
\label{sec:conclusions}

In this work we have characterized a lattice model of Dirac fermions
on a negatively curved background in presence of a local defect. The
unperturbed lattice model is the so-called {\em rainbow model}, which
is a free-fermionic chain with hoppings which decay exponentially from
the center. Its ground state presents linear growth of the entanglement, with an
entropy per site proportional to the inhomogeneity parameter $h$ in
the weak inhomogeneity regime, which is described by a geometrical
deformation of the free-fermionic conformal field theory, associated
to a hyperbolic space-time metric. The strong inhomogeneity limit is
described as a valence bond state with concentric bonds around the
center, as it can be established using the Dasgupta-Ma renormalization
group.

The presence of a defect in the center of the chain can induce an
entanglement transition in the strong inhomogeneity limit,
characterized by a rainbow phase of linear scaling of entanglement for
intermediate defect strengths, and two dimerized phases, with
alternate dimerizations in similarity with the SSH model. Further
hints of the transition are provided by the ground state energy, the
single-body orbitals, the energy gap (rescaled with the minimum
coupling) and two order parameters: the average dimerization and the
{\em rainbow order parameter}, which measures the average occupation
of the concentric bonds.

In the weak inhomogeneity limit, the transition gets blurred, and the
ground state always presents linear entanglement, with an entropy
per site that can be effectively described by a geometric deformation
of the entanglement entropy of a homogeneous fermionic chain with a
central defect. Analysis of the entanglement gap and the entanglement
Hamiltonian allow us to claim that the system behaves as a thermofield
double, as in the rainbow case, but with a dimerized Hamiltonian
instead of a homogeneous one.

Interestingly, the rainbow system presents non-trivial topological
properties when it is centered on a site instead of a link. In
presence of a defect, the ground state presents an interesting
coexistence of a symmetry-protected topological phase near the ends
and a dimerized region near the center, whose size grows as the defect
intensity goes to zero.

Our work opens up several interesting questions related to the
presence of geometric defects on the vacuum structure of a quantum
field theory. It is interesting to ask whether such a deep
modification of the entanglement properties can be found in other
cases.

\begin{acknowledgements}
We thank  A. Dauphin, J. I. Latorre,  A. Ludwig,  S. Ryu and  E. Tonni for conversations.  
We acknowledge the
Spanish government for financial support through grants
PGC2018-095862-B-C21 (NSSB and GS), PGC2018-094763-B-I00 (SNS),
QUITEMAD+ S2013/ICE-2801, SEV-2016-0597 of the ``Centro de Excelencia
Severo Ochoa'' Programme and the CSIC Research Platform on Quantum
Technologies PTI-001.
\end{acknowledgements}

\clearpage

\onecolumngrid

\appendix

\section{Dasgupta-Ma RG extension for free fermions}
\label{sec:rgextension}

In this Appendix we describe a generalization of the Dasgupta-Ma RG
for inhomogeneous free fermionic chains that can be applied to systems
that have an homogeneous subchain of $N=2L$ sites embedded. The
Hamiltonian $H_0$ that describes this subchain is given by

\begin{equation}
  \label{eq:hhomo}
  H_0=-\frac{J}{2}\sum_{i=1}^{N-1}c_i^\dagger c_{i+1}+c_{i+1}^\dagger c_i,
\end{equation}
and its interactions with the nearest neighbours
is given by $H_{lr}$

\begin{equation}
  H_{lr}=-J_lc_l^\dagger c_1- J_rc_N^\dagger c_r +\text{h.c.}.
\end{equation}
Assuming that $J_l\ll J$ and $\frac{J_l}{J_r}\approx1$ the whole
system can be study by means of degenerate  perturbation theory. The
ground state of $H_0$ is given in the previous Appendix
\ref{sec:single_body_modes}
$\ket{\psi_0}=\prod_{m=1}^L\hat{\phi}_{k_m}^\dagger\ket{0}$ with
energy
$E_0=\sum_{m=1}^{L}\epsilon_{k_m}=-2\sum_{m=1}^{L}\cos\left(\frac{m\pi}{N+1}\right)$. The
first order correction $\mel{\psi_0;l',r'}{H_{lr}}{\psi_0;l,r}$ (where
$\ket{\psi_i;l,r}=\ket{\psi_i}\otimes\ket{l}\otimes\ket{r}$)
vanishes. The matrix element $B_{l,r;l'r'}$ of the degenerate second
order contribution is given by:

\begin{equation}
  B_{l,r;l'r'}=\sum_{i\neq0}\sum_{l''r''}
  \frac{\mel{\psi_0;l,r}{H_{lr}}{\psi_i;l'',r''}
    \mel{\psi_i;l'',r''}{H_{lr}}{\psi_0;l',r'}}{E_0-E_i}
\end{equation}
Expanding this product and taking into  account that
$\sum_{l''r''}\ket{l'',r''}\bra{l'',r''}=\mathbb{I}$ we have:

\begin{eqnarray}
\nonumber
B_{l,r;l'r'}&=J_l^2\left(\mel{l,r}{c_l^\dagger c_l}{l',r'}\sum_{i=1}^N\dfrac{\mel{\psi_0}{c_1}{\psi_i}\mel{\psi_i}{c_1^\dagger}{\psi_0}+\mel{\psi_0}{c_1^\dagger}{\psi_i}\mel{\psi_i}{c_1}{\psi_0}}{\epsilon_{k_i}}-\sum_{i=1}^N\dfrac{\mel{\psi_0}{c_1^\dagger}{\psi_i}\mel{\psi_i}{c_1}{\psi_0}}{\epsilon_{k_i}}\right)+\\ \nonumber
&J_r^2\left(\mel{l,r}{c_r^\dagger c_r}{l',r'}\sum_{i=1}^N\dfrac{\mel{\psi_0}{c_N}{\psi_i}\mel{\psi_i}{c_N^\dagger}{\psi_0}+\mel{\psi_0}{c_N^\dagger}{\psi_i}\mel{\psi_i}{c_N}{\psi_0}}{\epsilon_{k_i}}-\sum_{i=1}^N\dfrac{\mel{\psi_0}{c_N^\dagger}{\psi_i}\mel{\psi_i}{c_N}{\psi_0}}{\epsilon_{k_i}}\right)+\\\nonumber
&J_lJ_r\left(\mel{l,r}{c_l^\dagger c_r}{l',r'}\sum_{i=1}^N\dfrac{\mel{\psi_0}{c_1}{\psi_i}\mel{\psi_i}{c_N^\dagger}{\psi_0}+\mel{\psi_0}{c_N^\dagger}{\psi_i}\mel{\psi_i}{c_1}{\psi_0}}{\epsilon_{k_i}}+\right.\\
&\left.\mel{l,r}{c_r^\dagger c_l}{l',r'}\sum_{i=1}^N\dfrac{\mel{\psi_0}{c_N}{\psi_i}\mel{\psi_i}{c_1^\dagger}{\psi_0}+\mel{\psi_0}{c_1^\dagger}{\psi_i}\mel{\psi_i}{c_N}{\psi_0}}{\epsilon_{k_i}}\right)
\end{eqnarray}

Where the non vanishing  contributions
are given by the excited states whose particle number differs by  one with
respect to $\ket{\psi_0}$:

\begin{eqnarray}
\mel{\psi_i}{c_i^\dagger}{\psi_0}\neq0\quad \text{if}\quad \ket{\psi_i}=\hat{\phi}_{k_i}\ket{\psi_0}, \quad E_i=E_0-\epsilon_{k_i},\\
\mel{\psi_i}{c_i}{\psi_0}\neq0\quad \text{if}\quad \ket{\psi_i}=\hat{\phi}_{k_i}^\dagger\ket{\psi_0}, \quad E_i=E_0+\epsilon_{k_i} \ . 
\end{eqnarray}
Given  that $c_i=\sum_{m=1}^Nf^*_{im}\hat{\phi}_{k_m}$ we
reach

\begin{eqnarray}
\nonumber
&B_{l,r;l'r'}=J_l^2\left(\mel{l,r}{c_l^\dagger c_l}{l',r'}\sum_{i=1}^N\dfrac{|f_{1m}|^2}{\epsilon_{k_m}}+\sum_{i=1}^L\dfrac{|f_{1m}|^2}{\epsilon_{k_m}}\right)+J_r^2\left(\mel{l,r}{c_r^\dagger c_r}{l',r'}\sum_{i=1}^N\dfrac{|f_{Nm}|^2}{\epsilon_{k_m}}+\sum_{i=1}^L\dfrac{|f_{Nm}|^2}{\epsilon_{k_m}}\right)+\\ \nonumber
&J_lJ_r\left(\mel{l,r}{c_l^\dagger c_r}{l',r'}\sum_{i=1}^N\dfrac{f_{1m}^*f_{mN}}{\epsilon_{k_m}}+\mel{l,r}{c_r^\dagger c_l}{l',r'}\sum_{i=1}^N\dfrac{f_{Nm}^*f_{m1}}{\epsilon_{k_m}}\right).
\end{eqnarray}

Now, particularizing for the functions Eq. \eqref{eq:hmodes} we obtain
that the renormalized Hamiltonian $B$ is (up to an additive constant):

\begin{equation}
  H_{eff}=-\frac{J_lJ_r}{J}\left(c_l^\dagger c_r+\text{h.c}\right),
\end{equation}
which is the expression used to renormalize the systems with strength
defects $\gamma=0$ and $\gamma=1$.


\section{The transition blocks: $\gamma=1$ and $\gamma=0$.}
\label{sec:single_body_modes}

In this appendix we derive  Eq. \eqref{eq:bond4c}. Consider an
homogeneous system of $N=2L$ sites with open boundary conditions (OBC)
whose Hamiltonian is given by Eq. \eqref{eq:hhomo}. Solving the time
independent Schrödinger equation
$H_0\ket{\phi_k}=\epsilon_k\ket{\phi_k}$ we arrive at

\begin{eqnarray}
\label{eq:hmodes}
\epsilon_{k_m}&=&-J\cos\left(\frac{m\pi}{N+1}\right)\\ \nonumber
\ket{\phi_{k_m}}&=&\hat{\phi}_{k_m}^\dagger\ket{0}=\sum_{i=1}^Nf^*_{mi}c_i^\dagger\ket{0},\quad\text{within }f^*_{mi}=\sqrt{\frac{2}{N+1}}\sin\left(\frac{m\pi i}{N+1}\right).
\end{eqnarray}
Taking  $N=4$, the many body ground state $\ket{\psi}$
at half filling is obtained by occupying the single body levels with
energies $-J\cos\frac{\pi}{5}$ and $-J\cos\frac{2\pi}{5}$.

\begin{equation}
  \ket{\psi_0}=d^\dagger\ket{0}=v^\dagger u^\dagger\ket{0}=
  \hat{\phi}_{k_2}^\dagger\hat{\phi}_{k_1}^\dagger\ket{0}.
\end{equation}


\section{Correlation matrices and entanglement entropy}	
\label{sec:corrmatrix}

The correlation matrices  $C$ for the ground states
Eqs. (\ref{eq:gsr})-(\ref{eq:gs1})
are given by 

\begin{eqnarray}
 C_{ij}&=& \ev{c^\dagger_ic_j}{GS} \nonumber
 = \sum_{m,m'=1}^L f_{i,m}f_{m',j}^*\ev{\hat{\phi}_m^\dagger\hat{\phi}_{m'}}{GS}\\
 &=& \sum_{m=1}^Lf_{i,m}f_{m,j}^*,
\label{eq:corr}
\end{eqnarray}
where we consider half filling and $\hat{\phi}_m$ are the fermionic
excitations of each system ($b_{i,j}$ Eq. \eqref{eq:bonding} and $d_i$
Eq. \eqref{eq:bond4c} in our case) and $f_{i,k}$ is a unitary matrix.

The EE entropy of a block $A_\ell$ is given by
\cite{Peschel.03}:

\begin{equation}
\label{eq:eecorr}
S(A_\ell)=-\sum_{k=1}^\ell \nu_k\log{\nu_k}+(1-\nu_k)\log{(1-\nu_k)},
\end{equation}
where the $\{\nu_k\}$ are the set of eigenvalues of the correlation matrix 
restricted to the block 
$A_\ell$. We shall  next describe  the correlation matrices as a function
of the  defect parameter $\gamma$. All the matrices are symmetric $C_{i,j}=C_{j,i}$
and present left-right symmetry $C_{i,j}=C_{N+1-j,N+1-i}$. All the
computations are done with $L$ even.

\begin{itemize}
\item $\gamma<0$:
  \begin{eqnarray}
    \begin{cases}
      C_{i,i}=\frac{1}{2},  \quad i=1, \dots, L\\
      C_{1,N}=-\frac{1}{2} , \\
      C_{2i,2i+1}=\frac{1}{2}, \quad i =1,\dots\frac{L}{2},
    \end{cases}
  \end{eqnarray}
\item $\gamma=0$:
  \begin{eqnarray}
    \begin{cases}
      C_{1,N}=-\frac{1}{2},\quad C_{L,L+1} = \frac{1}{2}\\
      C_{i,i}=\frac{1}{2},  \quad i=1, \dots, L\\
      C_{i,N+1-i}=(-1)^i\frac{1}{2\sqrt{5}}, \quad i=2, \dots, L-1\\
      C_{2i,2i+1}=\frac{1}{\sqrt{5}} , \quad i =1,\dots\frac{L}{2},
    \end{cases}
  \end{eqnarray}

\item $\gamma\in(0,1)$:
  \begin{eqnarray}
    C_{i,j}&=&\frac{1}{2}\delta_{i,i}+(-1)^{i}\frac{1}{2}\delta_{i,N+1-i} , 
  \end{eqnarray}
\item $\gamma=1$:
  \begin{eqnarray}
    \begin{cases}
      C_{i,i}&=\frac{1}{2} , \quad i=1, \dots, L\\
      C_{i,N+1-i}&=(-1)^i\frac{1}{2\sqrt{5}} ,  \quad i=1, \dots, L\\
      C_{2i-1,2i}&=\frac{1}{\sqrt{5}},  \quad i =1,\dots\frac{L}{2}
    \end{cases}
  \end{eqnarray}
\item $\gamma>1$:
  \begin{eqnarray}
    \begin{cases}
      C_{i,i}=\frac{1}{2},\quad & i =1, \dots, L\\
      C_{2i-1,2i}=\frac{1}{2} , \quad  & i =1,\dots\frac{L}{2}
    \end{cases}
  \end{eqnarray}
  
The correlation matrix of the four sites that are integrated
out in the same step whose ground state is given by
Eq. (\ref{eq:bond4c}) is

\begin{equation}
  C_4=\left(
  \begin{array}{cccc}
    \frac{1}{2} & \frac{1}{\sqrt{5}} & 0 & -\frac{1}{2\sqrt{5}} \\
    \frac{1}{\sqrt{5}} & \frac{1}{2} & \frac{1}{2\sqrt{5}} & 0 \\
    0 & \frac{1}{2\sqrt{5}} & \frac{1}{2} & \frac{1}{\sqrt{5}} \\
    -\frac{1}{2\sqrt{5}} & 0 & \frac{1}{\sqrt{5}} & \frac{1}{2} \\
  \end{array}
  \right).
\end{equation}
The most simple non trivial lateral block is

\begin{equation}
  A_2=\left(
  \begin{array}{cc}
    \frac{1}{2} & \frac{1}{\sqrt{5}} \\
    \frac{1}{\sqrt{5}} & \frac{1}{2} \\
  \end{array}
  \right),
\end{equation}
whose eigenvalues are $\nu_1=\frac{1}{10} \left(2 \sqrt{5}+5\right),\nu_2=\frac{1}{10} \left(5-2 \sqrt{5}\right)$. 
The value of $S_a$, given in  Eq. (\ref{eq:Sa}),  is obtained applying
Eq. (\ref{eq:eecorr}). Furthermore $S_b$, given in Eq. (\ref{eq:Sb},  is obtained from the central block

\begin{equation}
  B_1=\left(\begin{array}{cc}
    \frac{1}{2} & \frac{1}{2\sqrt{5}} \\
    \frac{1}{2\sqrt{5}} & \frac{1}{2} \\
  \end{array}
  \right),
\end{equation}
whose eigenvalues are $\nu_1=\frac{1}{10} \left(\sqrt{5}+5\right),\nu_2=\frac{1}{10} \left(5-\sqrt{5}\right)$. 
It can be shown  that larger central blocks have also these non trivial eigenvalues and the rest are 0 and 1 which don't contribute to the EE.
\end{itemize}


\section{Relation with Dirac equation with $\delta$ potential} 
\label{sec:relation_with_dirac_equation_with_}

Consider an inhomogeneous free-fermion chain with a central hopping
defect and bond centered symmetry described by the Hamiltonian:

\begin{equation}
\label{eq:homohamapendix}
H(\tau)= -\frac{\tau}{2} c_{-\frac{1}{2}}^\dagger c_{\frac{1}{2}}-
\frac{1}{2}\sum_{m=\frac{1}{2}}^{L-\frac{3}{2}}J_m
\(c^\dagger_{m}c_{m+1}+ c^\dagger_{-m}c_{-m+1}\).
\end{equation}
The single body spectrum is obtained by diagonalizing the hopping matrix. 
The eigenvalue equations at the center of the chain are

\begin{eqnarray}
  \alpha\phi_{-\frac{3}{2}}+\tau\phi_{\frac{1}{2}}&=&\epsilon\phi_{-\frac{1}{2}}\\
  \label{eq:centerII}
  \tau\phi_{-\frac{1}{2}}+\alpha\phi_{\frac{3}{2}}&=&\epsilon\phi_{\frac{1}{2}}
\end{eqnarray}
where $\epsilon$ is the single body energy and $\phi_m$ is the
amplitude associated with the fermionic operator $c_m$ and
$J_{\frac{1}{2}}=\alpha$. The expansion of the local operators $c_m$
in terms of its right and left moving components around the Fermi
points  (see Eq. \eqref{eq:continuum_limit}) leads to the equations:

\begin{eqnarray}
  \label{eq:center1}
  \tau\left(\psi_L^{II}-i\psi_R^{II}\right)&=
  \left(-i\epsilon+\alpha\right)\psi_L^{I}+\left(\epsilon-i\alpha\right)\psi_R^{I}\\
  \label{eq:center2}
  \tau\left(\psi_L^{I}+i\psi_R^{I}\right)&=
  \left(i\epsilon+\alpha\right)\psi_L^{II}+\left(\epsilon+i\alpha\right)\psi_R^{II},
\end{eqnarray}
with

\begin{eqnarray}
  \lim_{a\to0}\psi_{L,R}\(-\frac{3}{2}a\right)&
  =&\lim_{a\to0}\psi_{L,R}\left(-\frac{1}{2}a\)=\psi_{L,R}^I,\\
  \lim_{a\to0}\psi_{L,R}\(\frac{3}{2}a\)&
  =&\lim_{a\to0}\psi_{L,R}\(\frac{1}{2}a\)=\psi_{L,R}^{II},
\end{eqnarray}
Solving for $\psi_L^I$ in Eq. (\ref{eq:center1}) and putting it into
Eq. (\ref{eq:center2}) we have:

\begin{equation}
  \label{eq:psir}
  \psi_R^I=\frac{1}{2\alpha\tau}\(i(\tau^2-\alpha^2-\epsilon^2)\psi_L^{II}+
  (\tau^2-(\epsilon+i\alpha)^2)\psi_R^{II}\).
\end{equation}
Inserting this expression into Eq. (\ref{eq:center1}) we arrive at

\begin{equation}
  \psi_L^I=\frac{1}{2\alpha\tau}
  \((\tau^2+(\alpha+i\epsilon)^2)\psi_L^{II}-i(\tau^2-\alpha^2+\epsilon^2)\psi_R^{II}\).
\end{equation}
We can express these two equations as $\psi^I=T\psi^{II}$, where $T$
is a transfer matrix:

\begin{equation}
  \begin{pmatrix}
    \psi_L^I\\
    \psi_R^I
  \end{pmatrix}
  =\frac{1}{2\alpha\tau}
  \begin{pmatrix}
    \tau^2+(\alpha+i\epsilon)^2 & -i(\tau^2-\alpha^2+\epsilon^2)\\
    i(\tau^2-\alpha^2-\epsilon^2)& \tau^2-(\epsilon+i\alpha)^2
  \end{pmatrix}
  \begin{pmatrix}
    \psi_L^{II}\\
    \psi_R^{II}
  \end{pmatrix}.
\end{equation}

Furthermore, at half filling we have that $\epsilon\xrightarrow[L \to
  \infty]{}0$ and the transfer matrix simplifies to:

\begin{equation}
  T=\frac{1}{2\alpha\tau}\begin{pmatrix}\tau^2+\alpha^2 & -i(\tau^2-\alpha^2)\\
    i(\tau^2-\alpha^2)& \tau^2+\alpha^2\end{pmatrix}.
\end{equation}

Note that this can be also written as:

\begin{equation}
  T=\frac{1}{2\tilde{\tau}}\begin{pmatrix}\tilde{\tau}^2+1 & -i(\tilde{\tau}^2-1)\\
    i(\tilde{\tau}^2-1)& \tilde{\tau}^2+1\end{pmatrix},
\end{equation}
where $\tilde{\tau}=\frac{\tau}{\alpha}$. Substituting $\tau=e^{-h\gamma}$ and $\alpha=e^{-\frac{h}{2}}$ we have the expression Eq. \eqref{eq:transfergamma}.

\section{Details of the RG applied to the SCS system}
\label{sec:detailsscs}

In this Appendix we derive the ground state of the Hamiltonian 

\begin{equation}
  H_N(h,\gamma)_\scs=-\frac{1}{2}\sum_{m=1}^N
  J_m c_m^\dagger c_{m+1} + \text{h.c.},\quad\text{with}\quad J_m = \begin{cases}
  e^{-h(\left|m - (L + \frac{1}{2})\right| - \frac{1}{2})} & \text{if
  } m\neq L,L+1,\\
  e^{ -h\gamma }  & \text{if } m\in\{L,L+1\} \ . 
\end{cases}
  \label{eq:hamscs}
\end{equation}

We use the RG scheme explained in the main text. 
There are three cases to be considered: 

\bigskip

(1.-) Case $\gamma<1$. The couplings present
a {\em double tie} at the center, so that the dominant interaction
involves the three central sites, $L$, $L+1$ and $L+2$. The
Dasgupta-Ma prescription Eq. \eqref{eq:sdrg} and the sum rule
Eq. \eqref{eq:teorema} are not valid in this situation. We must
perform a first order perturbation approach to renormalize three
fermionic sites into an effective site (see Appendix A of
\cite{Samos.19}), leading to a system with $N-2$ sites. The next RG
step involves the effective fermion mode created on the previous one
and its two nearest neighbours. Iterating this procedure one obtains
the GS:
    
\begin{equation}
  \ket{GS}_{\gamma<1}=(g^\xi_L)^\dagger
  \prod_{m=1}^{L-2}(f^{s_m}_m)^\dagger (g^+_0)^\dagger\ket{0},
\end{equation}
where $s_m=(-1)^m$,  $\xi=(-1)^\frac{L}{2}$ and

\begin{eqnarray} 
  g^\pm_m &=&\frac{1}{\sqrt{2}} (c_{L+1}+b^\pm_{L-2m,L+2(m+1)}),\\
  g^\pm_L &=&\frac{1}{\sqrt{2}} (c_{1}+b^\pm_{2,2L})\\
  f^\pm_n &=&\frac{1}{\sqrt{2}}(b^\pm_{L+1-n,L+1+n}+b^\pm_{L-n,L+2+n}),
\end{eqnarray}
with $b^\pm_{i,j}$ defined in Eq. \eqref{eq:bonding}. 

\vspace{0.5cm} 

(2.-) Case  $\gamma=1$.  The system presents a {\em quadruple
  tie} at the center. The five central sites involved are
renormalized into an effective site on a system with $N-4$ sites. At
this point the situation is equivalent to the $\gamma<1$ case and
further RG steps are the same as the ones discussed in the previous
item.

\vspace{0.5cm}

\begin{figure}
\centering
  \includegraphics[width=8cm]{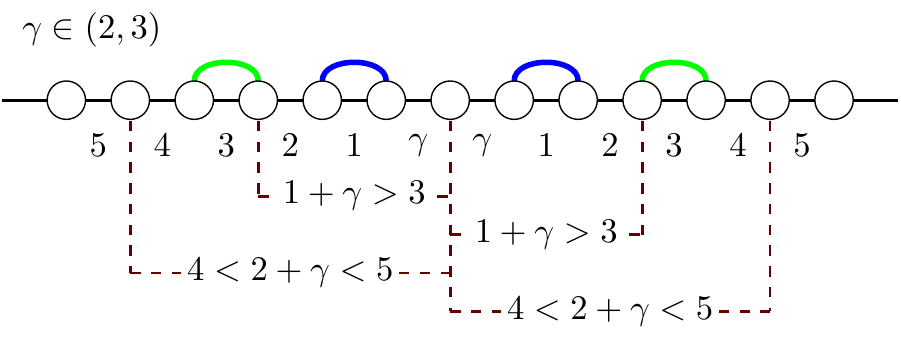}
  \caption{RG procedure for a system $H_N(h,\gamma)_\scs$ with $h\gg1$
    and $\gamma\in(2,3)$. The system admits two Dasgupta-Ma RG steps,
    above links with log-coupling $1$, and after those the
    renormalized system shows a {\em double tie} of lowest
    log-couplings, with the two central log-couplings equal to
    $1+\gamma$ (which must be larger than $3$).  At this moment, we
    apply the same RG procedure than for $\gamma<1$.}
  \label{fig:rg_scs}
\end{figure}

(3.-) Case $\gamma>1$.   In this situation, the
dominant interactions are two non-consecutive log-couplings 1, which
couple respectively the sites $L-1$ and $L$, and the sites $L+2$ and
$L+3$. Although the sum rule Eq. \eqref{eq:teorema} is not valid, the
Dasgupta-Ma Eq. \eqref{eq:sdrg} can be applied sequentially twice,
yielding two fermionic excitations with the same energy and parity,
$b^+_{L-1,L}$ and $b^+_{L+2,L+3}$, and leading to a effective system
whose Hamiltonian is $H_{N-4}(h,1+\gamma)_\scs$. The next decimation
step is not univocal:

\begin{itemize}
\item If $\gamma\in (1,2)$ the dominant interaction involves the three
  central sites $L-2$,$L$ and $L+4$ of the original chain and the
  situation is the same as the one described originally for
  $\gamma<1$, with the double tie (1.).
\item If $\gamma=2$ there is a quadruple tie, same as (2.).
\item If $\gamma\in (2,3)$ the dominant interactions are the two links with
  log-coupling 3, which is similar to the situation described in item
  (3.). At the end of this step there are two fermionic excitations
  more, $b^+_{L-3,L-2}$ and $b^+_{L+4,L+5}$, and the Hamiltonian of
  the decimated system is $H_{N-8}(h,2+\gamma)_\scs$. We show in
  Fig. \ref{fig:rg_scs} this situation.
\end{itemize}

Note that unless $\gamma>L-1$ the RG flows towards the double tie
situation and, if $\gamma\in\mathbb{N}$, the decimated system of the
$\gamma$-th step will present a quadruple tie. Hence, the GS is:
\begin{equation}
\ket{GS}_{\gamma>1}=(g_L^-)^\dagger
  \prod_{k=1}^{L-2(1+\floor{\gamma})}(f^{\eta_k}_{k+2\floor{\gamma}})^\dagger
  (g^-_{\floor{\gamma}})^\dagger
  \prod_{m=1}^{\floor{\gamma}}(b^+_{L-2(m-1),L-2(m-1)-1})^\dagger
  (b^+_{L+2m,L+2m+1})^\dagger\ket{0},
  \end{equation}

where $\floor{}$ is the floor function, $\eta_k=(-1)^{k+1}$,
$\chi_k=(-1)^{k-\floor{\gamma}}$.\\


\end{document}